\documentclass[10pt,journal]{IEEEtran}
\usepackage[T1]{fontenc}
\usepackage[utf8]{inputenc}
\usepackage{bm}
\usepackage{amsmath}
\usepackage{amssymb}
\usepackage{mathtools}
\usepackage{hyperref}
\usepackage[nameinlink]{cleveref}
\usepackage{color}
\usepackage{nidanfloat}
\usepackage{ifthen}

\usepackage{graphicx}
\graphicspath{{figures/}}

\usepackage[backend=biber, style=ieee, natbib=true, mincitenames=1, maxcitenames=2, giveninits=true, uniquename=init]{biblatex}

\makeatletter
\DeclareFieldFormat{eprint:openreview}{%
  OpenReview\addcolon\space
  \ifhyperref
    {\href{https://openreview.net/forum?id=#1}{\nolinkurl{#1}}}
    {\nolinkurl{#1}}}
\makeatother
\makeatletter
\newcounter{IEEE@bibentries}
\renewcommand\IEEEtriggeratref[1]{%
  \renewbibmacro{finentry}{%
    \stepcounter{IEEE@bibentries}%
    \ifthenelse{\equal{\value{IEEE@bibentries}}{#1}}
    {\finentry\@IEEEtriggercmd}
    {\finentry}%
  }%
}
\makeatother

\usepackage{tikz}
\usetikzlibrary{backgrounds, calc, chains, fit, matrix, positioning, shadows, shapes, circuits.ee}
\tikzset{input/.style={}}
\tikzset{output/.style={}}
\tikzset{op/.style={circle, draw, thick, fill=black!10, minimum size=2.5ex, inner sep=0ex}}
\tikzset{filter/.style={rectangle, draw, thick, fill=black!10, minimum size=3.5ex, inner sep=1ex}}
\tikzset{nn/.style={trapezium, trapezium angle=80, draw, thick, fill=black!10, inner sep=1ex}}
\tikzset{branch/.style={circle, draw, thick, fill=black, minimum size=.5ex, inner sep=0ex}}
\tikzset{tensor/.style={rectangle, draw, thick, fill=white, minimum size=2em, double copy shadow={shadow xshift=.5ex,shadow yshift=-.5ex}}}
\tikzset{image/.style={rectangle, draw, thick, fill=white, minimum size=2em}}
\tikzset{>=direction ee}

\definecolor{gblue}{HTML}{1f77b4}
\definecolor{ggreen}{HTML}{2ca02c}

\DeclareMathOperator{\softplus}{softplus}

\DeclareMathOperator*{\argmin}{arg\,min}

\DeclarePairedDelimiterX{\divergence}[2]{[}{]}{#1\;\delimsize\|\;#2}
\newcommand{\KL}[2]{D_\text{KL}\divergence{#1}{#2}}
\newcommand{\E}{\operatorname{\mathbb E}}
\newcommand{\R}{\operatorname{\mathbb R}}

\newcommand{\Z}{\operatorname{\mathbb Z}}
\newcommand{\D}{\;\mathrm{d}}
\newcommand{\T}{\top}
\newcommand{\const}{\mathrm{const}}

\newcommand\comment[1]{}

\IEEEtriggeratref{96}

\begin{document}
%
\title{Nonlinear Transform Coding}

\author{
\IEEEauthorblockN{
Johannes Ballé, Philip A.\ Chou, David Minnen, Saurabh Singh,\\Nick Johnston, Eirikur Agustsson, Sung Jin Hwang, George Toderici}

\IEEEauthorblockA{
Google Research\\
Mountain View, CA 94043, USA\\
\{jballe, philchou, dminnen, saurabhsingh, nickj, eirikur, sjhwang, gtoderici\}@google.com}
}
\date{}


\maketitle

\begin{abstract}
We review a class of methods that can be collected under the name nonlinear transform coding (NTC), which over the past few years have become competitive with the best linear transform codecs for images, and have superseded them in terms of rate--distortion performance under established perceptual quality metrics such as MS-SSIM. We assess the empirical rate--distortion performance of NTC with the help of simple example sources, for which the optimal performance of a vector quantizer is easier to estimate than with natural data sources. To this end, we introduce a novel variant of entropy-constrained vector quantization. We provide an analysis of various forms of stochastic optimization techniques for NTC models; review architectures of transforms based on artificial neural networks, as well as learned entropy models; and provide a direct comparison of a number of methods to parameterize the rate--distortion trade-off of nonlinear transforms, introducing a simplified one.
\end{abstract}


\begin{figure*}
  \includegraphics[width=.49\linewidth]{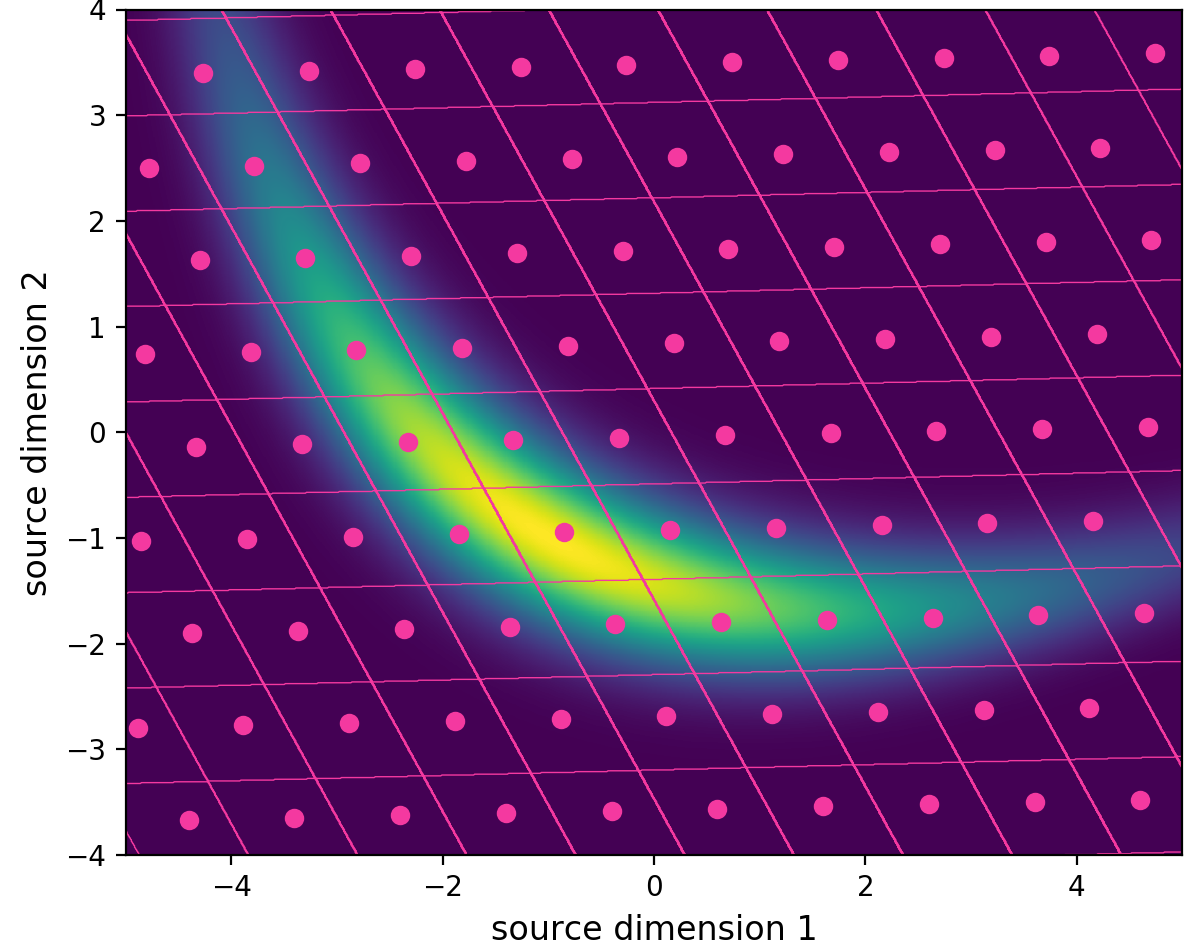}\hfill%
  \includegraphics[width=.49\linewidth]{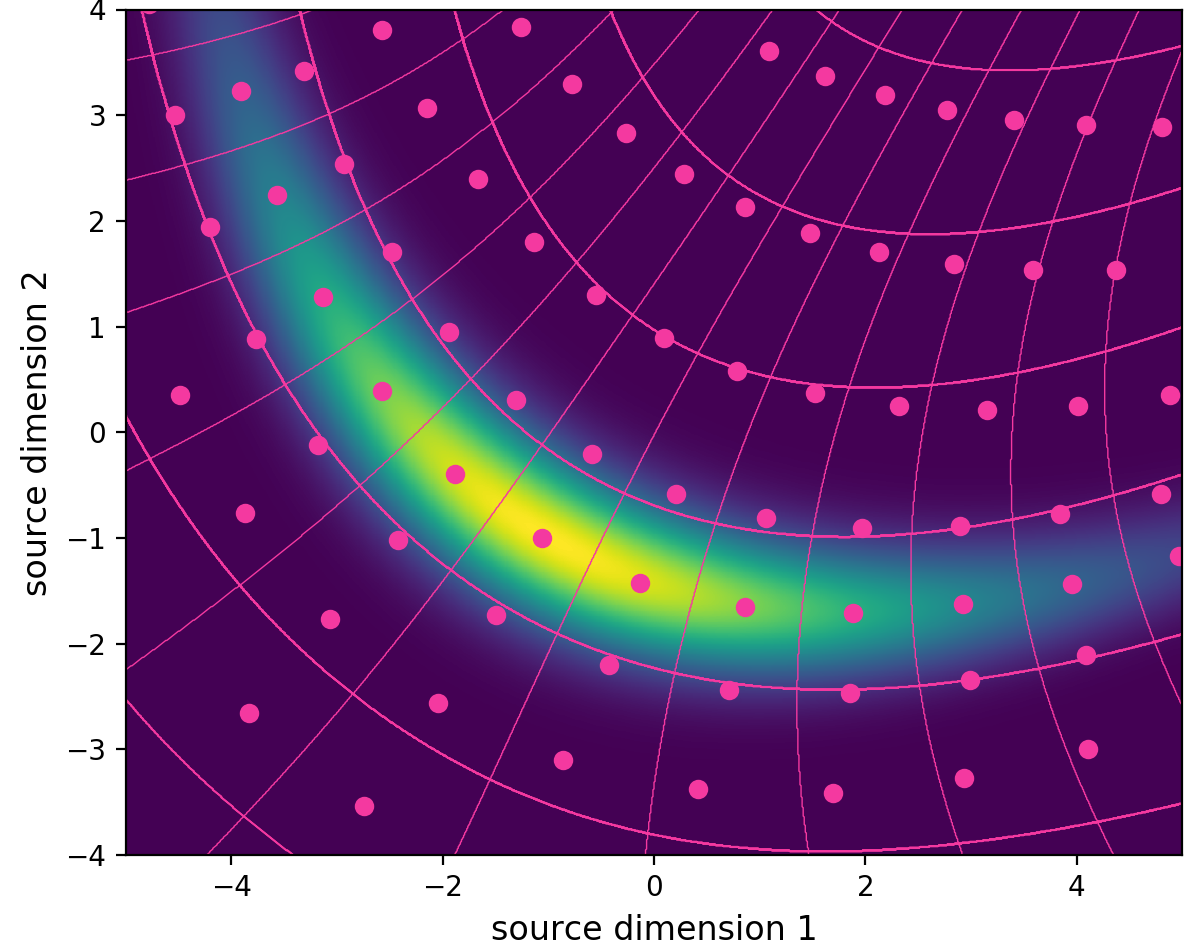}
  \caption{Linear transform code (left), and nonlinear transform code (right) of a banana-shaped source distribution, both obtained by empirically minimizing the rate--distortion Lagrangian (\cref{eq:loss_noisy}). Lines represent quantization bin boundaries, while dots indicate code vectors. While LTC is limited to lattice quantization, NTC can more closely adapt to the source, leading to better compression performance (RD results in \cref{fig:rd_laplace_banana}; details in \cref{sec:ntc}).}
  \label{fig:ntc_banana}
\end{figure*}

\section{Introduction}
There is no end in sight for the world's reliance on multimedia communication. Digital devices have been increasingly permeating our daily lives, and with them comes the need to store, send, and receive images and audio ever more efficiently. Almost universally, transform coding (TC) has been the method of choice for compressing this type of data source.

In his 2001 article for IEEE Signal Processing Magazine \citep{Go01}, Vivek Goyal attributed the success of TC to a divide-and-conquer paradigm: the practical benefit of TC is that it separates the task of decorrelating a source, from coding it. Any source can be optimally compressed in theory using vector quantization (VQ) \citep{GeGr92}. However, in general, VQ quickly becomes computationally infeasible for sources of more than a handful dimensions, mainly because the codebook of reproduction vectors, as well as the computational complexity of the search for the best reproduction of the source vector grow exponentially with the number of dimensions. TC simplifies quantization and coding by first mapping the source vector into a latent space via a decorrelating invertible transform, such as the Karhunen--Loève Transform (KLT), and then separately quantizing and coding each of the latent dimensions.

Much of the theory surrounding TC is based on an implicit or explicit assumption that the source is jointly Gaussian, because this assumption allows for closed-form solutions. If the source is Gaussian, all that is needed to make the latent dimensions independent is decorrelation. When speaking of TC, it is almost always assumed that the transforms are linear, even if the source is far from Gaussian. As an example, consider the banana-shaped distribution in \cref{fig:ntc_banana}: While linear transform coding (LTC) is limited to lattice quantization, nonlinear transform coding (NTC) can more closely adapt to the source, leading to better compression performance.

Until a few years ago, one of the fundamental constraints in designing transform codes was that determining nonlinear transforms with desirable properties, such as improved independence between latent dimensions, is a difficult problem for high-dimensional sources.
As a result, not much practical research had been conducted in directly using nonlinear transforms for compression.
However, this premise has changed with the recent resurgence of artificial neural networks (ANNs).
It is well known that, with the right set of parameters, ANNs can approximate arbitrary functions \citep{LeLiPiSc93}.
It turns out that in combination with stochastic optimization methods, such as stochastic gradient descent (SGD), and massively parallel computational hardware, a nearly universal set of tools for function approximation has emerged.
These tools have also been used in the context of data compression \citep{ToOMHwViMi16,BaLaSi17,ThShCuHu17,ToViJoHwMi17,RiBo17,AgMeTsCaTi17}.
Even though these methods were developed from scratch, they have rapidly become competitive with modern conventional compression methods such as HEVC \citep{HEVC}, which are the culmination of decades of incremental engineering efforts.
This demonstrates, as it has in other fields, the flexibility and ease of prototyping that universal function approximation brings over designing methods manually, and the power of developing methods in a data-driven fashion.

This paper reviews some of the recent developments in data-driven lossy compression; in particular, we focus on a class of methods that can be collectively called \emph{nonlinear transform coding} (NTC), providing insights into its capabilities and challenges. We assess the empirical rate--distortion (RD) performance of NTC with the help of simple example sources: the Laplace source and the two-dimensional distribution of \cref{fig:ntc_banana}. To this end, we introduce a novel variant of entropy-constrained vector quantization (ECVQ) algorithm \citep{chou1989entropy}. Further, we provide insights into various forms of optimization techniques for NTC models and review ANN-based transform architectures, as well as entropy modeling for NTC. A further contribution of this paper is to provide a direct comparison of a number of methods to parameterize the RD trade-off, and to introduce a simplified method.

In the next section, we first review stochastic gradient optimization of the RD Lagrangian, a necessary tool for optimizing ANNs for lossy compression. We introduce \emph{variational} ECVQ, illustrating this type of optimization. VECVQ also serves as a baseline to evaluate NTC in the subsequent section. In that section, we discuss various approaches for approximating the gradient of the RD objective and review ANN architectures. \Cref{sec:entropy_modeling} reviews entropy modeling via learned forward and backward adaptation, and illustrates its performance gains on image compression. \Cref{sec:multirate} compares several ways of parameterizing the transforms to continuously traverse the RD curve with a single set of transforms. The last two sections discuss connections to related work and conclude the paper, respectively.

\section{Stochastic rate--distortion optimization}
Consider the following lossy compression scenario. Alice is drawing vectors $\bm x \in \R^N$ from some data source, whose probability density function we denote $p_\textrm{source}$. Here, Alice is concerned with compressing each vector into a bit sequence, communicating this sequence to Bob, who then uses the information to reconstruct an approximation to $\bm x$. Each possible vector $\bm x$ is approximated using a codevector $\bm c_k \in C$, where $C = \{ \bm c_k \in \R^N \mid 0 \le k < K \}$ is called the codebook. Once the codevector index $k = e(\bm x)$ for a given $\bm x$ is determined using the encoder $e(\cdot)$, Alice subjects it to lossless entropy coding, such as Huffman coding or arithmetic coding, which yields a bit sequence of nominal length $s(k)$. In what follows, we'll assume that the performance of this entropy coding method is optimized to closely approximate the theoretical limit, i.e., that Alice and Bob share an estimate of the marginal probability distribution of $k$, also called an entropy model, $P(k)$, and that $s(k) \approx -\log P(k)$. To the extent that $P(k)$ approximates $M(k)=\E_{\bm x \sim p_\textrm{source}} \delta(k, e(\bm x))$, the true marginal distribution of $k$ (where $\delta$ denotes the Kronecker delta function), $s(k)$ is close to optimal, since codes of length $-\log M(k)$ would achieve the lowest possible average rate, the entropy of $k$. Since Alice and Bob also share knowledge of the codebook, Bob can decode the index $k$ and finally look up the reconstructed vector $\bm c_k$.

\begin{figure}
  \includegraphics[width=\linewidth]{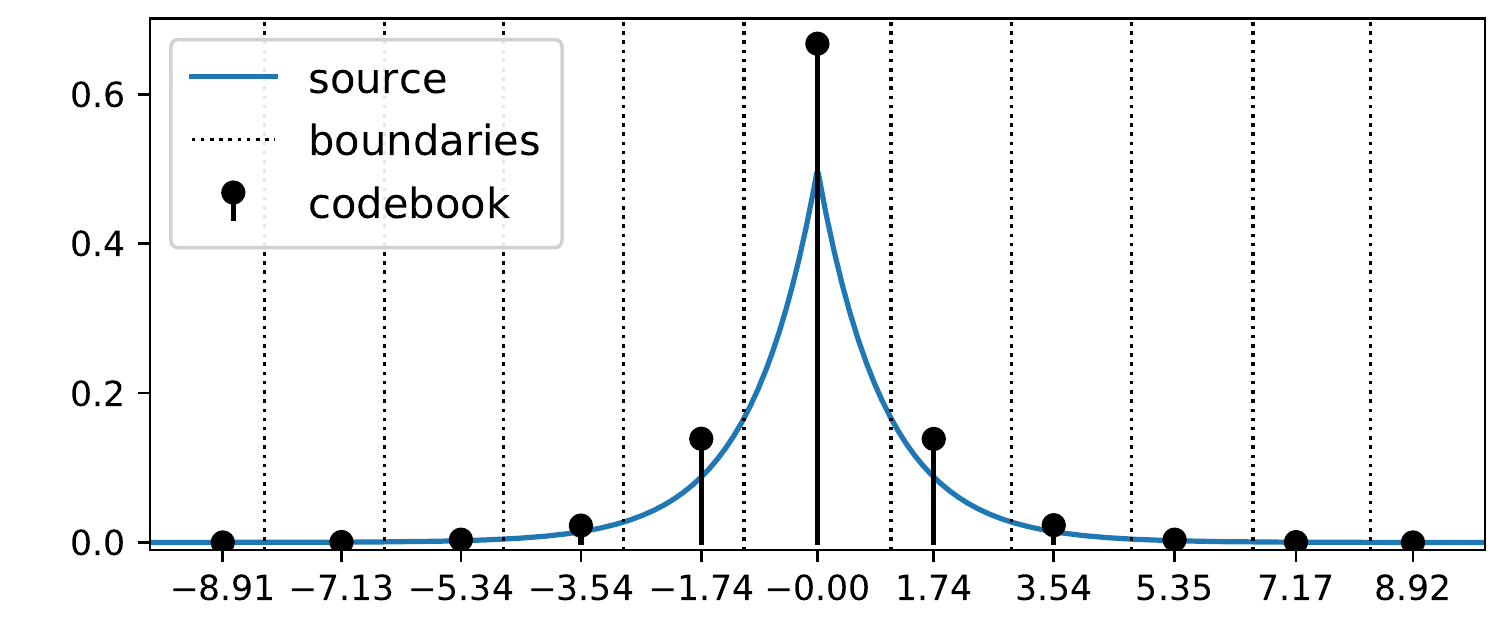}

  \includegraphics[width=\linewidth]{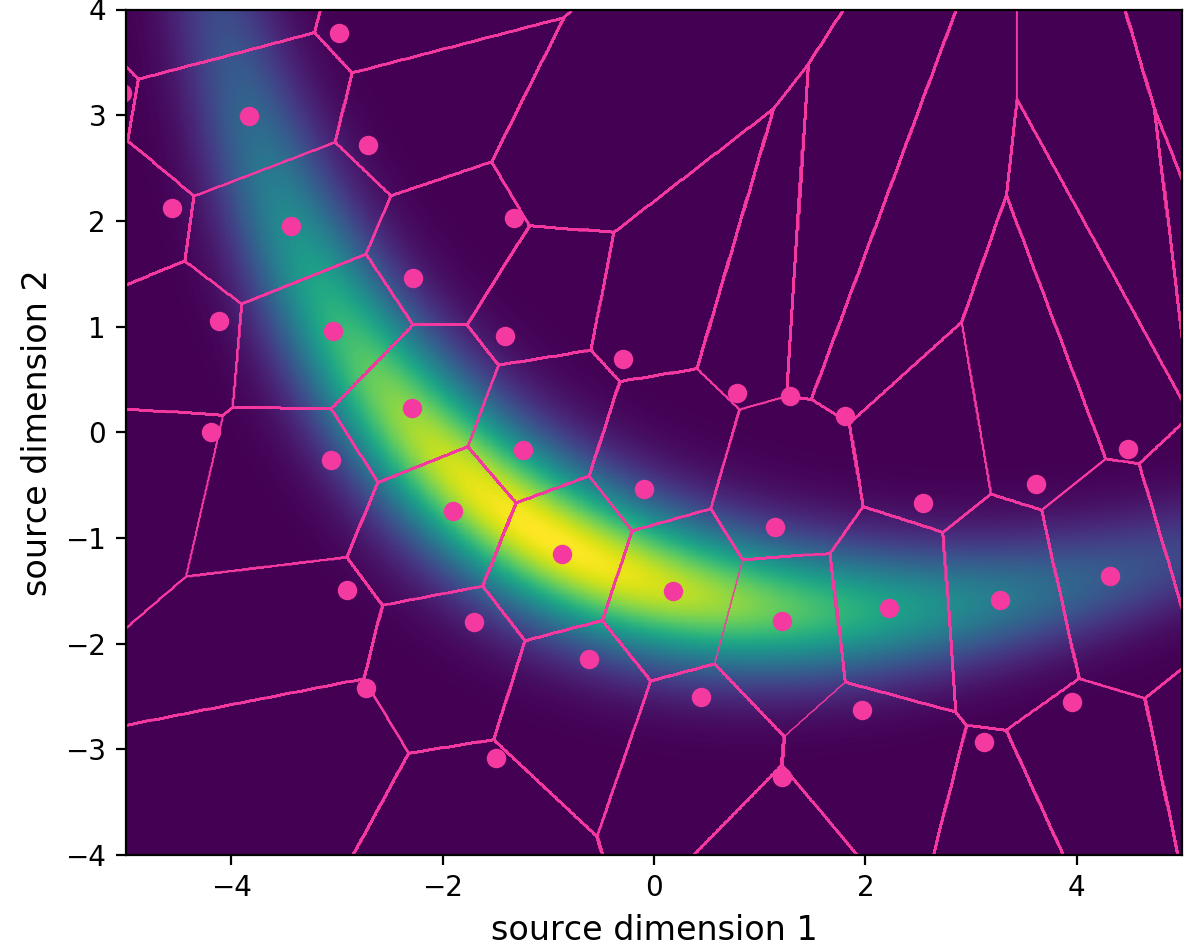}
  \caption{Top: A near-optimal entropy-constrained scalar quantizer of a standard Laplacian source, found using the VECVQ algorithm (\cref{eq:loss_vq}). Bottom: an entropy-constrained vector quantizer of a banana-shaped source, found using the same algorithm.}
  \label{fig:vecvq}
\end{figure}

To optimize the efficiency of this scheme, Alice and Bob seek to simultaneously minimize the cross entropy of the index under the entropy model (the rate) as well as the distortion between $\bm x$ and the reconstructed vector, quantified by some distortion measure $d$:
\begin{equation}
L = \E_{\bm x \sim p_\textrm{source}} \bigl[ -\log P(k) + \lambda \, d(\bm x, \bm c_k) \bigr],
\label{eq:loss}
\end{equation}
with $k = e(\bm x)$ as determined by the encoder $e(\cdot)$, choosing a codebook index for each possible source vector.
Many authors formulate this as a minimization problem over one of the terms given a hard constraint on the other \citep{CoTh06}.
In this paper, we consider the Lagrangian relaxation of the distortion-constrained problem, with the Lagrange multiplier $\lambda$ on the distortion term determining the trade-off between rate and distortion.

The top panel of \cref{fig:vecvq} illustrates a lossy compression method for a simple, one-dimensional Laplacian source, optimized for squared error distortion (i.e., $d(\bm x, \bm c) = \| \bm x - \bm c \|_2^2$). The source distribution is plotted in blue. The codebook vectors are represented by the horizontal locations of the black stalks, while the height of each stalk is proportional to the likelihood of that code vector under the entropy model $P$. Dotted lines delineate the quantization bins, i.e., the intervals for which all source values get mapped to a given codebook value (the one within the respective interval). For Laplacian sources, the minimizer of $L$ has been studied by \citet{Su96}. It is characterized by equal-width quantization bins and equidistant code vectors, except for the center bin (coinciding with the mode of the source distribution); both characteristic features are present in the figure up to small deviations. The bottom panel of the same figure visualizes a vector quantizer for a banana-shaped source distribution. The boundaries between quantization bins are shown as pink lines, while the code vectors are rendered as discs. Note the presence of hexagon-like bins, which are a feature of optimal VQ for squared-error distortions.

\subsection{Variational entropy-constrained vector quantization}
To generate both of the results in \cref{fig:vecvq}, we used a novel algorithm for entropy-constrained vector quantization based on directly minimizing \cref{eq:loss}.
To begin, without loss of generality, we parameterize the entropy model as
\begin{equation}
P(k) = \frac{e^{a_k}}{\sum_{j=0}^{K-1} e^{a_j}}.
\label{eq:categorical_prior}
\end{equation}
Then, denoting model parameters $\Theta = \{a_k, \bm c_k \mid 0 \le k < K\}$, we define the sample loss
\begin{equation}
    \ell_\Theta(k, \bm x) = -\log P(k) + \lambda \, d(\bm x, \bm c_k) 
\end{equation}
and the encoder function
\begin{equation}
    e_\Theta(\bm x) = \argmin_k \ell_\Theta(k, \bm x),
\end{equation}
where we have made explicit their dependence on the parameters $\Theta$.
We express \cref{eq:loss} as
\begin{equation}
    L_\text{VQ} = \E_{\bm x} \ell_\Theta(e_\Theta(\bm x), \bm x) = \E_{\bm x} \min_k \ell_\Theta(k, \bm x),
    \label{eq:loss_vq}
\end{equation}
which we now wish to minimize over $\Theta$ using stochastic gradient descent (SGD).
SGD relies on a Monte Carlo approximation of the expectation, and the fact that expectations and derivatives are both linear operators, whose order can be exchanged.
Thus
\begin{align}
  \frac \partial {\partial \Theta} L_\text{VQ}
  &= \E_{\bm x} \frac \partial {\partial \Theta} \min_k \ell_\Theta(k, \bm x),
\intertext{which can be approximated by the sample expectation}
  \frac \partial {\partial \Theta} L_\text{VQ}
  &\approx \frac 1 B \sum_{\{\bm x_b \sim p_\text{source} \mid 0 \le b < B\}} \frac {\partial \ell_\Theta(k_b, \bm x_b)} {\partial \Theta},
\label{eq:sgd}
\end{align}
with $k_b = e_\Theta(\bm x_b)$.
This is an unbiased estimator of the derivative of $L_\text{VQ}$ based on averaging the derivatives of $\ell$ over a batch of $B$ source vector samples.

Minimization of $L_{\text VQ}$ will fit the entropy model to the marginal distribution of $k$, $M(k) = \E_{\bm x \sim p_\textrm{source}} \delta(k, e(\bm x))$, as well as adjust the codebook vectors to minimize distortion. To see this, add and subtract the expected negative log likelihood of $k$ under the marginal to \cref{eq:loss_vq}:
\begin{equation}
L_\text{VQ} = \KL M P + \E_{\bm x} \bigl[-\log M(k) + \lambda \, d(\bm x, \bm c_k) \bigr].
\label{eq:vq_variational}
\end{equation}
Since the second term is constant wrt. the parameters of $P$ almost everywhere, minimizing $L_\text{VQ}$ results in fitting $P$ to $M$ by minimizing their Kullback--Leibler (KL) divergence. Similarly, since the first term is constant wrt. the codebook almost everywhere, each $\bm c_k$ is adjusted to minimize the distortion between it and all source vectors getting mapped to $k$. Note that since the KL divergence is non-negative, $L_\text{VQ}$ can be interpreted as an upper bound on the second term, which is the rate--distortion objective for the optimal choice of entropy model. This can be likened to variational Bayesian inference, in which a \emph{variational} approximation ($P$) to an unobserved true distribution ($M$) is found by minimizing an upper bound on the true objective. We therefore name this method variational entropy-constrained vector quantization (VECVQ).\footnote{The VECVQ algorithm inherits from two methods.  The first is the ECVQ algorithm of \citet{Berger72,FavardinM84,chou1989entropy}, which minimizes \cref{eq:loss} using a clustering algorithm instead of gradient descent.  The second is the online $K$-means algorithm of \citet{bottou1995convergence}, which minimizes the distortion part of \cref{eq:loss}, $\E_{\bm x} \bigl[d(\bm x, \bm c_k) \bigr]$, using gradient descent.  Both ECVQ and the online $K$-means algorithms derive in turn from the generalized Lloyd algorithm \citep{Lloyd82,Max60,macqueen1967some,LindeBG80}.} Note that, since $P$ as defined in \cref{eq:categorical_prior} can represent arbitrary distributions, the variational approximation here is capable of recovering the true marginal, i.e., the KL divergence can converge to zero.

\begin{figure*}
  \includegraphics[width=0.49\linewidth]{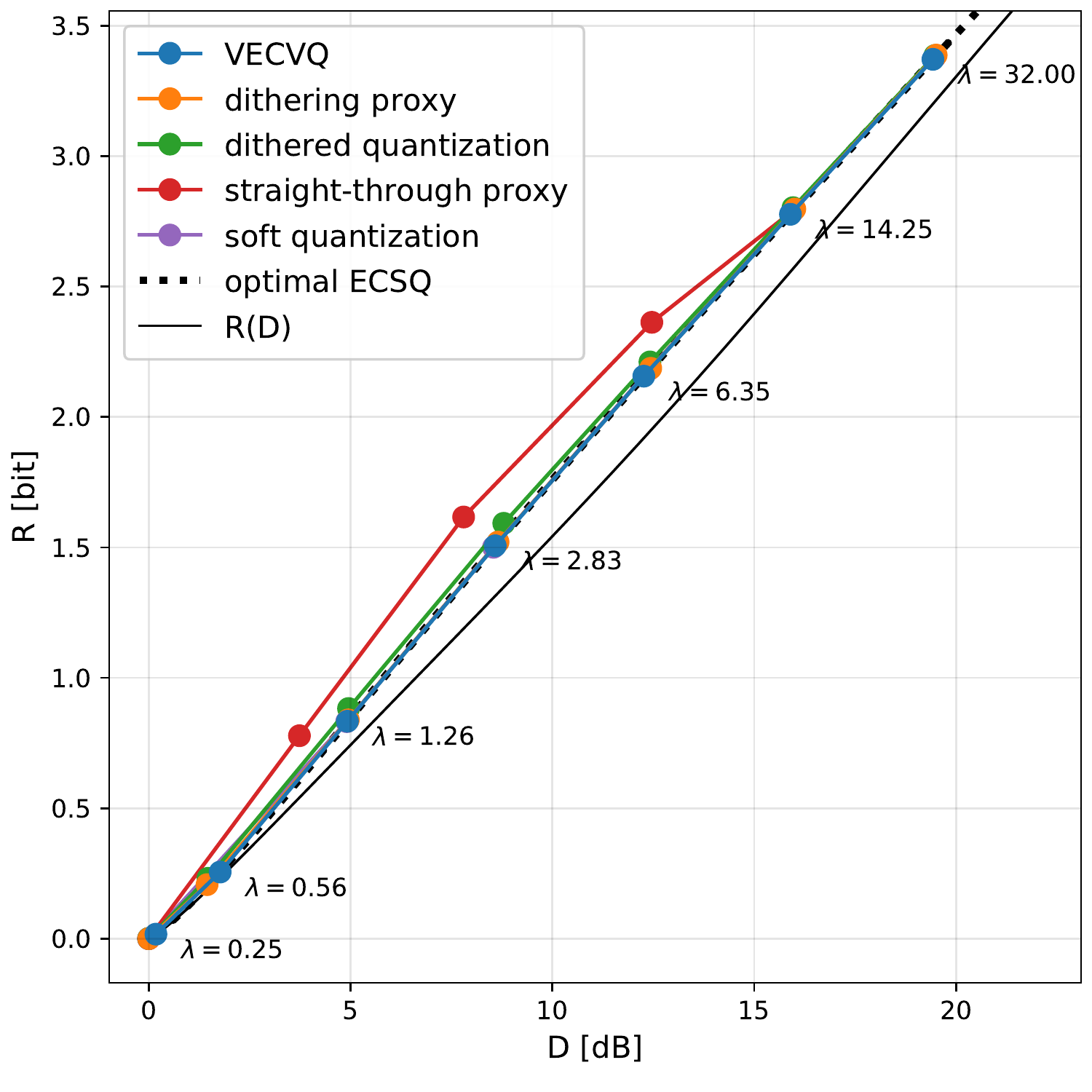}\hfill%
  \includegraphics[width=0.49\linewidth]{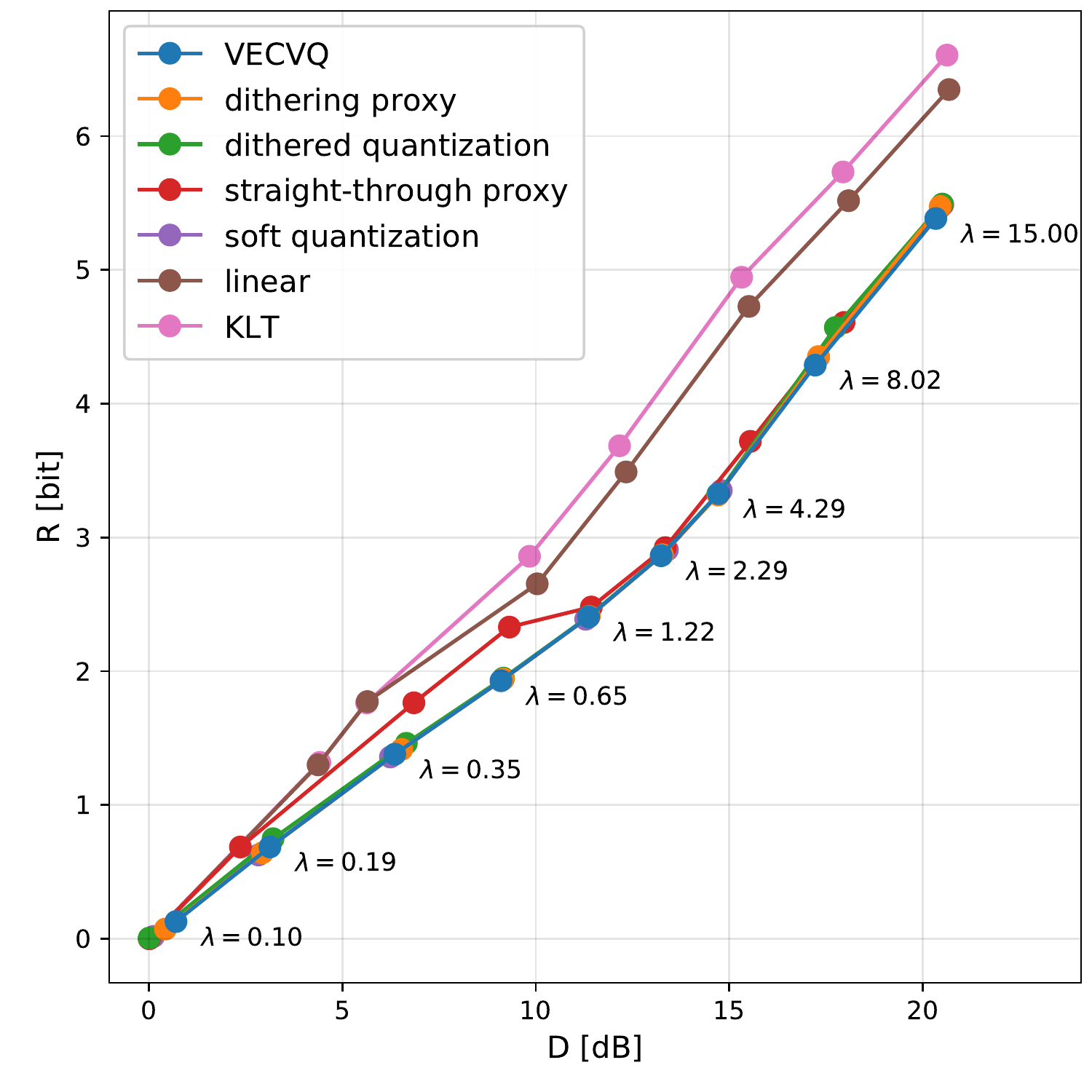}
  \caption{Left: rate--distortion performance of different quantizers for standard Laplace source. Both VECVQ and NTC with optimized offset (“dithering proxy”) recover the optimal entropy-constrained scalar quantizer established by \citet{Su96}. NTC with randomized offset (“dithered quantization”) is slightly suboptimal at lower rates, as predicted by theory. The NTC trained with the straight-through proxy is unstable at low rates. Using the dithering proxy with explicit soft quantization recovers the optimal quantizer as well. $R(D)$ indicates the information-theoretic rate--distortion function, $R(D)=\inf_{p(\hat x \mid x)}\{I(x;\hat x)$ s.t.\ $\E[d(x,\hat x)]\leq D\}$ (achievable only in the limit of large blocksizes, not with a scalar quantizer). Right: rate--distortion performance of different quantizers for banana source. NTC closely matches the performance of VECVQ; the straight-through variant diverges at low rates. The linear TC trained for the same objective performs significantly worse. Constraining it to the KLT is not necessarily optimal, as pointed out by \citet{Go01}.}
  \label{fig:rd_laplace_banana}
\end{figure*}

In the left panel of \cref{fig:rd_laplace_banana}, we plot the operational rate--distortion function of the optimal entropy-constrained scalar quantizer due to \citet{Su96}, as well as the empirical rate--distortion function of the VECVQ algorithm for the same Laplace source. The plot shows that the algorithm recovers the theoretical optimum. Since it is constrained only by the size of the codebook, we can use the algorithm as an empirical lower bound on the rate--distortion objective of more constrained compression methods, such as nonlinear transform coding, even for source distributions for which no theoretical optimum is presently known. As an example, consider the more complex banana distribution in the right panel: the nonlinear transform coders trained with the dithering proxy (to be discussed in the next section) perform ever so slightly worse than VECVQ.

\section{Nonlinear transform coding}
\label{sec:ntc}
It is easy to modify \cref{eq:loss} to accommodate nonlinear transform coding.
Rather than explicitly enumerating the codebook vectors, we consider mapping the source vectors into a latent space $\R^M$ and back via a pair of transforms.
Quantization and compression takes place in this latent space.
Specifically, we define the analysis transform as a parametric function $\bm y = g_a(\bm x)$, implemented by a neural network with parameters $\bm \phi$, and the synthesis transform as a function $\bm{\tilde x} = g_s(\bm{\tilde y})$, with parameters $\bm \theta$. We can write the rate--distortion objective as:
\begin{equation}
L_\text{NTC} = \E_{\bm x} \bigl[-\log P\bigl(\lfloor g_a(\bm x) \rceil\bigr) + 
\lambda \, d\bigl(\bm x, g_s(\lfloor g_a(\bm x) \rceil)\bigr)\bigr],
\label{eq:loss_ntc}
\end{equation}
where $\lfloor \cdot \rceil$ denotes uniform scalar quantization (rounding to integers).
$P$ is now a probability distribution over a space of integer vectors, which take the role of the codebook index $k$ in \cref{eq:loss}.

\begin{figure}
  \includegraphics[width=\linewidth]{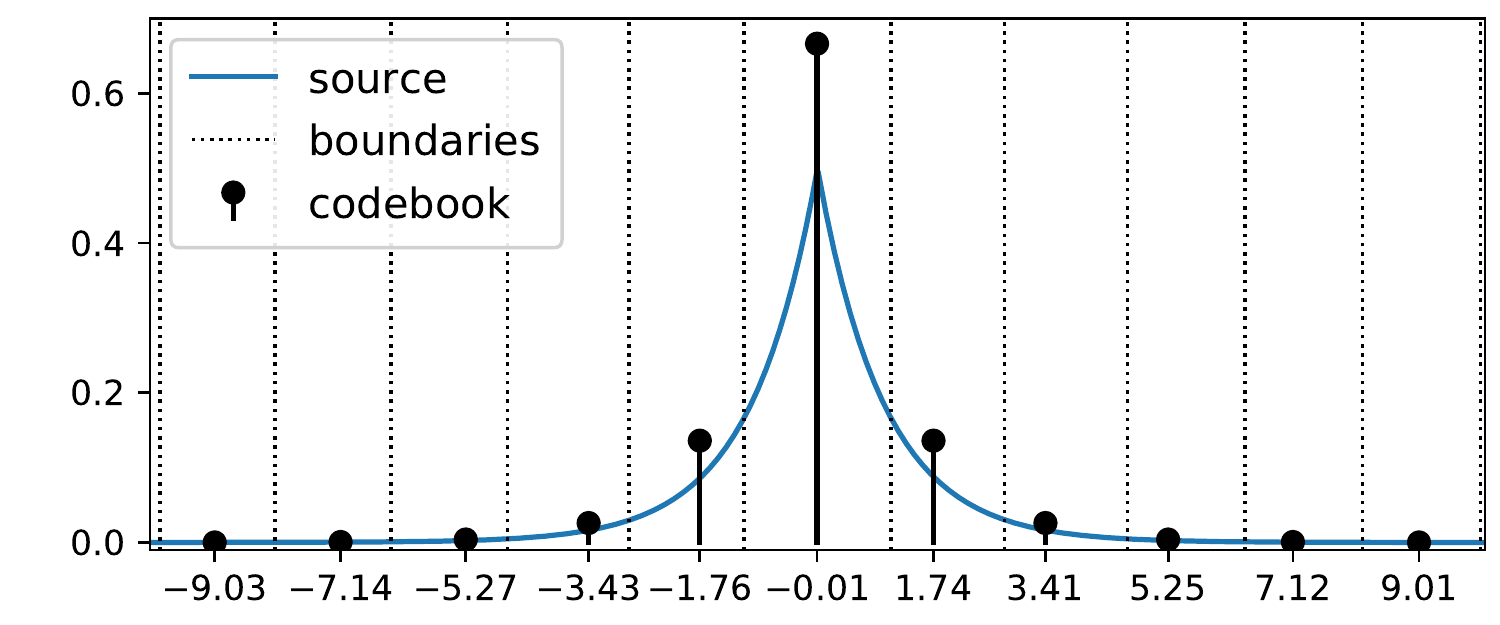}

  \includegraphics[width=\linewidth]{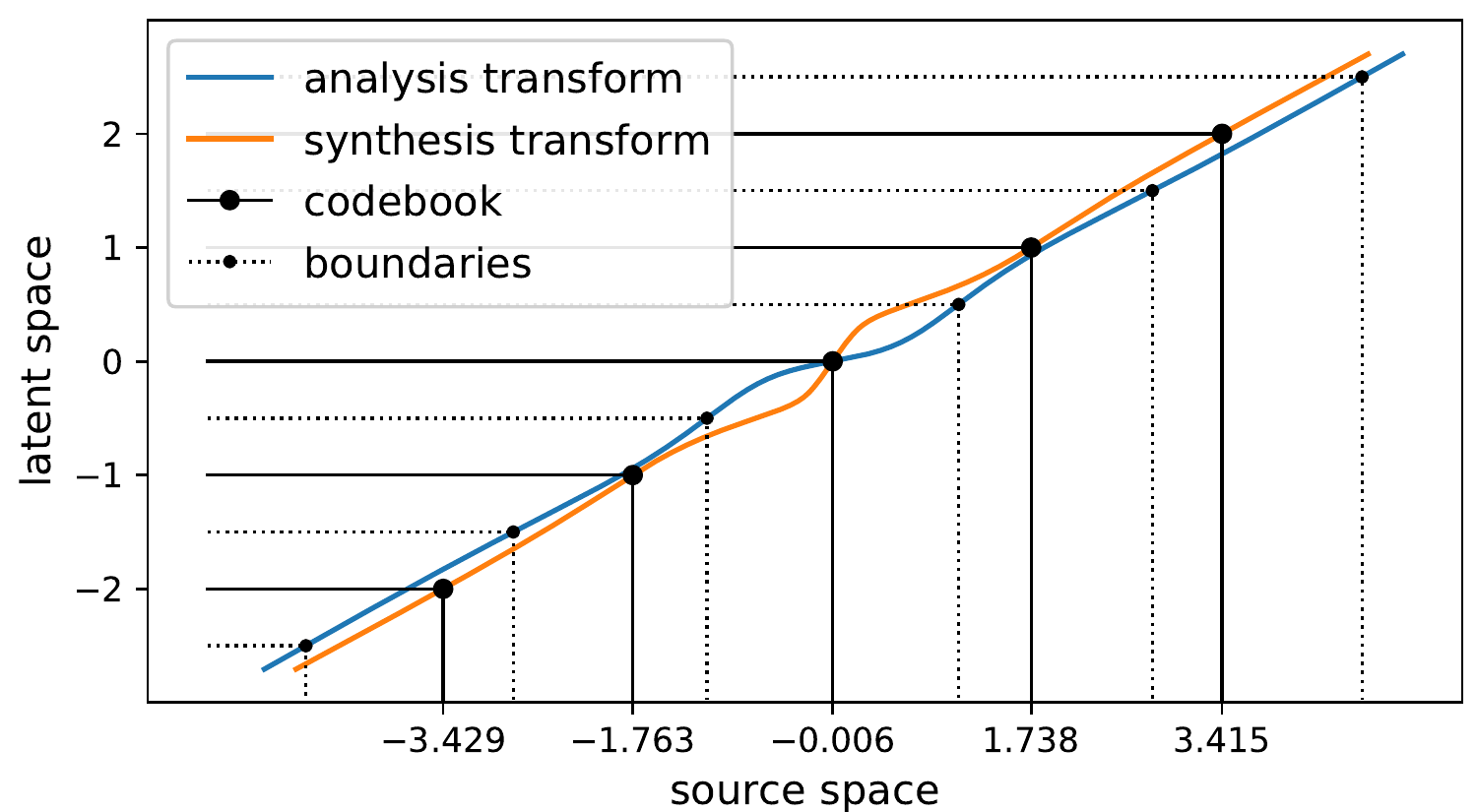}
  \caption{A near-optimal nonlinear transform code of a standard Laplacian source, obtained by minimizing the dithering rate--distortion proxy (\cref{eq:loss_noisy}).}
  \label{fig:ntc_laplacian}
\end{figure}

As an example, consider the nonlinear transform code illustrated in \cref{fig:ntc_laplacian}. Again, we plot the effective codebook vectors and quantization boundaries on top of the source distribution. However, unlike the example in \cref{fig:vecvq}, this quantization scheme is defined indirectly via the analysis and synthesis transforms, as illustrated in the bottom panel of \cref{fig:ntc_laplacian}. The analysis transform maps the space of source values to the latent space (blue curve). In this space, uniform quantization is applied, rounding values between half-integers to full integers. These integer values are then mapped back into the source space using the synthesis transform (orange curve).

There are a few key observations here: First, the analysis transform determines the effective quantization bins. In particular, its intersections with the dotted lines, corresponding to half-integers in the latent space, give rise to quantization bins of varying size in the source space. Second, the synthesis transform determines the effective codebook vectors. Notably, the full behavior of the synthesis transform as determined by the optimization procedure does not matter -- only its values at integer locations are relevant. Third, since the transforms are not constrained to be exact inverses of each other, using uniform quantization in the latent space is sufficient to enable codebook vectors to be located anywhere in the corresponding quantization bins (technically, even outside of it). Nonlinear transform coding thus generalizes companding \citep{Bennett48,Gersho79}, which permits implementing non-uniform quantization using uniform quantizers; with nonlinear transforms, the quantization step size can be fixed to one without loss of generality (parameterizing the model for different rate--distortion trade-offs is discussed in \cref{sec:multirate}).

\Cref{fig:ntc_banana} illustrates LTC and NTC of the banana distribution also shown in \cref{fig:vecvq} (bottom), and evaluated in \cref{fig:rd_laplace_banana} (right).
Both the linear and nonlinear transform codes are designed to minimize \cref{eq:loss_ntc} under their respective constraints.
Because the linear transform coder is constrained to affine transformations, the method effectively amounts to a lattice quantizer in the source space (left panel).
Note that the linear transform is not orthogonal and hence is \emph{not} the KLT, as would be optimal if the source distribution were Gaussian.

The nonlinear transform coder has more flexibility, and can adapt the shapes of its quantization bins to better fit the source distribution (right panel). Note that in both cases, since invertibility of the transforms is not enforced, codebook vectors do not necessarily appear in consistent locations relative to their bins. Their optimal locations, for squared error distortions, are at the conditional mean of their respective cells. Both coders reflect this by shifting the codebook vectors closer to the high-probability regions of the source distribution. This may come at the expense of reconstruction accuracy in low-probability regions -- in the nonlinear example, some low-probability codebook vectors even lie outside of their respective bins -- because the behavior of the method in these regions often does not contribute much to the overall objective. Another reason for this trade-off may be limitations in the parameterizations of the transforms. This is further examined in \cref{sec:transforms}.

\begin{figure*}
  \includegraphics[width=.49\linewidth]{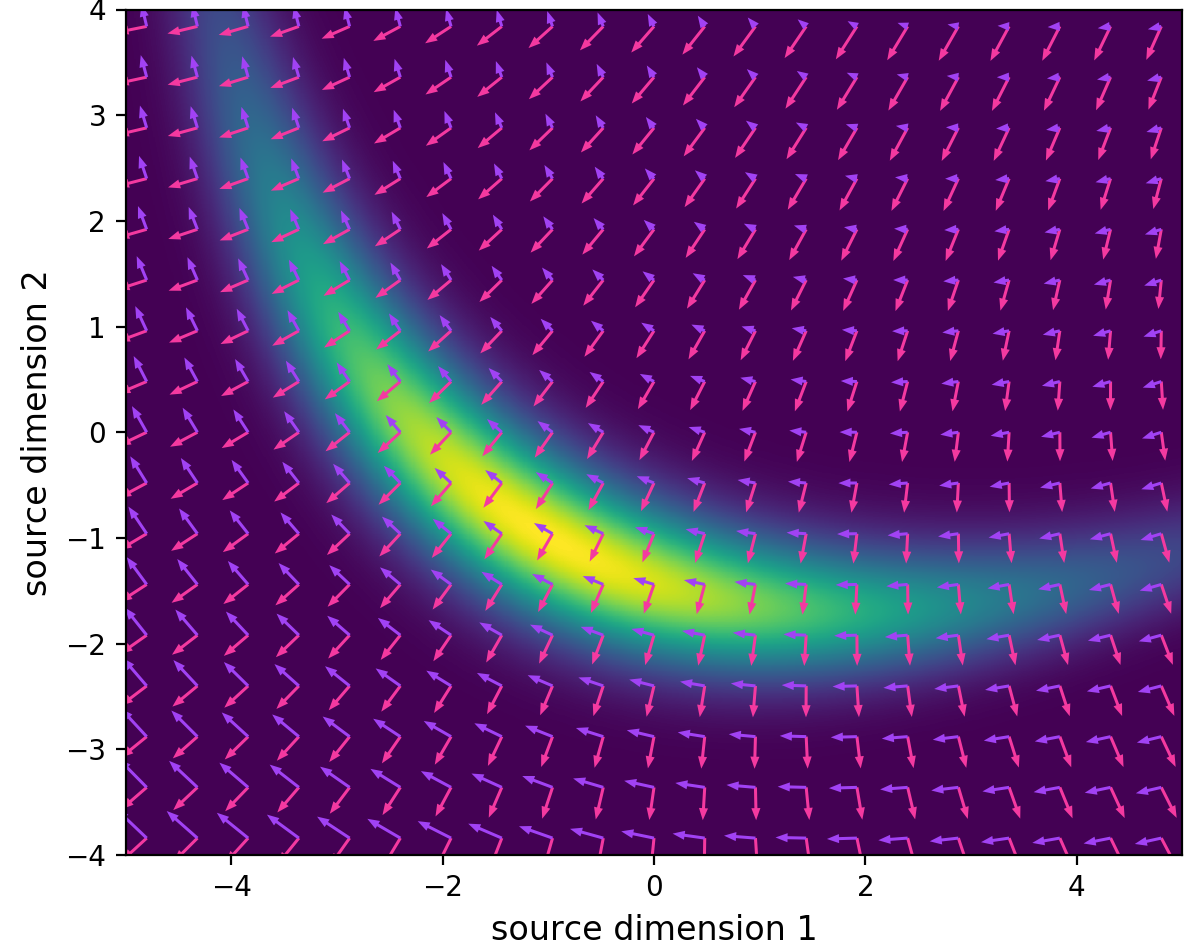}\hfill%
  \includegraphics[width=.49\linewidth]{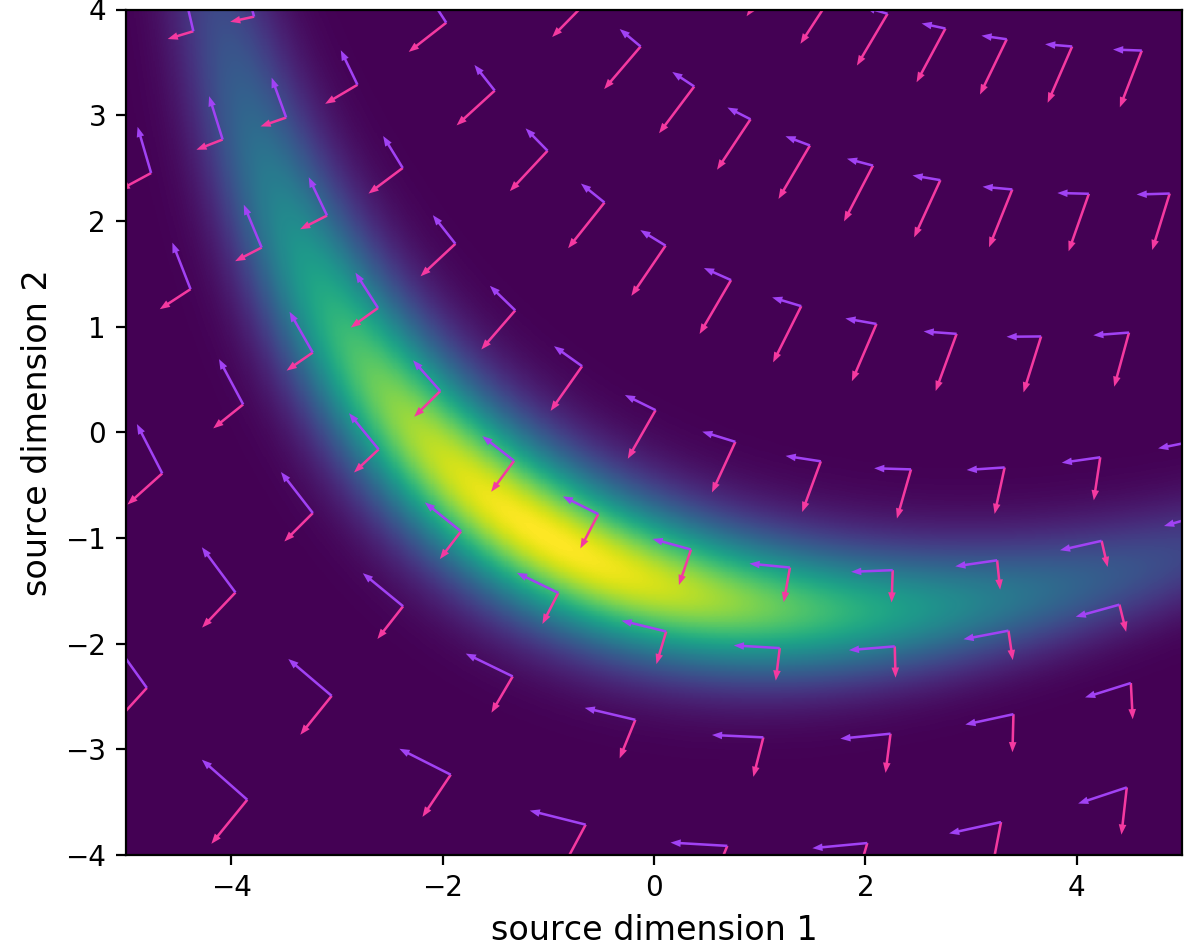}
  \caption{Location-dependent Jacobian matrices $\nabla g_a(\bm x)$ (left; arrows visualize local Jacobian inverse) and $\nabla g_s(\bm y)$ (right; arrows visualize local Jacobian) of a model optimized for squared error distortion on the banana source. The transforms form a local orthogonalization of the source density.}
  \label{fig:jacobians}
\end{figure*}

Although the optimized nonlinear analysis and synthesis transforms illustrated in \cref{fig:ntc_banana} (right) are globally nonlinear, they are of course differentiable, and hence can be viewed as locally approximately linear.
We find that in high-probability regions of the data distribution, they locally resemble KLTs in that they are approximately orthogonalized (\cref{fig:jacobians}).
Specifically, at each point $\bm x$, the columns of the inverse Jacobian matrix of the analysis transform appear approximately orthogonal to each other, as shown in the left panel of the figure. 
Likewise, the columns of the Jacobian matrix of the synthesis transform are approximately orthogonal, as shown in the right panel.
In the neighborhood of some point $\bm x_0$, $(\bm y - \bm y_0) \approx \nabla g_a(\bm x_0) \cdot (\bm x - \bm x_0)$ approximates a linear orthogonal analysis transform, and $(\bm{\tilde x} - \bm{\tilde x}_0) \approx \nabla g_s(\bm y_0) \cdot (\bm y - \bm y_0)$ approximates its inverse, where $\bm y_0=g_a(\bm x_0)$ and $\bm{\tilde x}_0=g_s(\bm y_0)$.

While more can be said about the local properties of the analysis and synthesis transforms (see supplement), for now, let us consider the transforms as “black boxes” that simply serve to approximate the optimal transforms.

\subsection{Optimization and proxy rate--distortion loss}
Note that in the VECVQ loss given in \cref{eq:loss_vq}, the encoder function is defined by exhaustively minimizing over all possible codes.
As such, it can be folded into a minimum over the sample loss $\ell_\Theta$, which is differentiable with respect to almost all $\bm x$.
If we were to choose another encoder function with trainable parameters of its own, we would not be able to obtain a gradient of the loss function with respect to them that is useful for SGD.
The gradient would be zero for almost all $\bm x$, because $e$ is integer valued.
This problem also appears in \cref{eq:loss_ntc} due to the quantizer.
Derivatives of the loss with respect to any parameter of the analysis transform are zero almost everywhere.
However, when employing dithered quantization (i.e., randomizing the quantization offset) \citep{Sc64}, this problem can be avoided \citep{BaLaSi16a}.

Consider uniformly sampling one random quantization offset per latent dimension $\bm o \in [-\frac 1 2, \frac 1 2)^M$, and formulating the following loss function as an expectation over it:
\begin{multline}
\E_{\bm o} L_{\text{NTC}, \bm o} = \E_{\bm x, \bm o} \Bigl[-\log P\bigl( \lfloor g_a(\bm x) - \bm o \rceil; \bm o \bigr) \\
+ \lambda \, d\bigl(\bm x, g_s(\lfloor g_a(\bm x) - \bm o \rceil + \bm o)\bigr)\Bigr],
\label{eq:loss_dither}
\end{multline}
where $L_{\text{NTC}, \bm o}$ is the loss for a given offset $\bm o$, and $P(\cdot; \bm o)$ is an entropy model conditioned on $\bm o$. (Note that all else being equal, the marginal distribution of the quantized latents changes with the offset.) This loss function is differentiable with respect to the parameters of $g_a$. To see this, let us consider both terms separately. For the rate term, we have
\begin{align}
&-\E_{\bm x, \bm o} \log P\bigl( \lfloor g_a(\bm x) - \bm o \rceil; \bm o \bigr) \notag \\
= &-\E_{\bm x, \bm o} \sum_{\bm k \in \Z^M} \delta\bigl( \lfloor g_a(\bm x) - \bm o \rceil = \bm k \bigr) \log P(\bm k; \bm o) \notag \\
= &-\E_{\bm x} \idotsint_{-\frac 1 2}^{\frac 1 2} \D \bm o \sum_{\bm k \in \Z^M} \notag \\
&\hspace{6em} \delta\bigl( \| g_a(\bm x) - \bm o - \bm k \|_\infty \le \tfrac 1 2 \bigr) \log P(\bm k; \bm o) \notag \\
= &-\E_{\bm x} \idotsint_{-\infty}^\infty \D \bm v \, \delta\bigl( \| g_a(\bm x) - \bm v \|_\infty \le \tfrac 1 2 \bigr) \log p(\bm v) \notag \\
= &-\E_{\bm x} \idotsint_{-\frac 1 2}^{\frac 1 2} \D \bm u \log p\bigl(g_a(\bm x) + \bm u\bigr) \notag \\
= &-\E_{\bm x, \bm u} \log p\bigl(g_a(\bm x) + \bm u\bigr),
\end{align}
where $\delta$ is the Kronecker delta function, we define $\bm v = \bm k + \bm o$ and $p(\bm v) = P(\lfloor \bm v \rceil; \bm v - \lfloor \bm v \rceil)$, and consider $\bm u \in [-\frac 1 2, \frac 1 2)^M$ uniformly distributed. It is easy to show that $p$ is non-negative and integrates to one, and hence represents a probability density function. We can thus interpret $p$ as a continuous equivalent of an entropy model for the “noisy” latents $g_a(\bm x) + \bm u$. For the distortion term, we have
\begin{multline}
\E_{\bm x, \bm o} d\bigl(\bm x, g_s(\lfloor g_a(\bm x) - \bm o \rceil + \bm o)\bigr) =\\
\E_{\bm x, \bm u} d\bigl(\bm x, g_s(g_a(\bm x) + \bm u)\bigr),
\end{multline}
since dithered quantization and additive uniform noise have the same marginal distribution (i.e., integrating out $\bm o$ and $\bm u$ is equivalent). For a proof, refer to \citet{Sc64}. 

Hence, denoting $\bm{\tilde y} = g_a(\bm x) + \bm u$, we can now write
\begin{equation}
\E_{\bm o} L_{\text{NTC}, \bm o} = \E_{\bm x, \bm u} \bigl[-\log p(\bm{\tilde y}) + \lambda \, d\bigl(\bm x, g_s(\bm{\tilde y})\bigr)\bigr],
\label{eq:loss_noisy}
\end{equation}
which can be directly minimized via SGD as in \cref{eq:sgd}.

Analogously to \cref{eq:vq_variational}, we can interpret \cref{eq:loss_noisy} as a variational upper bound on the true marginal:
\begin{multline}
\E_{\bm o} L_{\text{NTC}, \bm o} = \KL m p \\
+ \E_{\bm x, \bm u} \bigl[-\log m(\bm{\tilde y}) + \lambda \, d\bigl(\bm x, g_s(\bm{\tilde y})\bigr)\bigr],
\end{multline}
where $m(\bm v) = \E_{\bm x, \bm u} \delta(\bm v, g_a(\bm x) + \bm u)$ is the marginal distribution of the noisy latents.\footnote{This dithering objective of rate--distortion optimized nonlinear transform coding is equivalent to $\beta$-variational autoencoders \citep{BaLaSi17,HiMaPaBuGl17,AlPoFiDiSa18}, up to choice of parameterized distributions.} As such, minimizing \cref{eq:loss_noisy} results in fitting the continuous entropy model $p$ to the marginal. Note that, unlike in the case of VECVQ, the KL divergence may not converge to zero, as for high-dimensional source distributions such as images, the entropy model will generally not be capable of representing the marginal accurately. \Cref{sec:entropy_modeling} talks about this in more detail.

\begin{figure}
  \includegraphics[width=\linewidth]{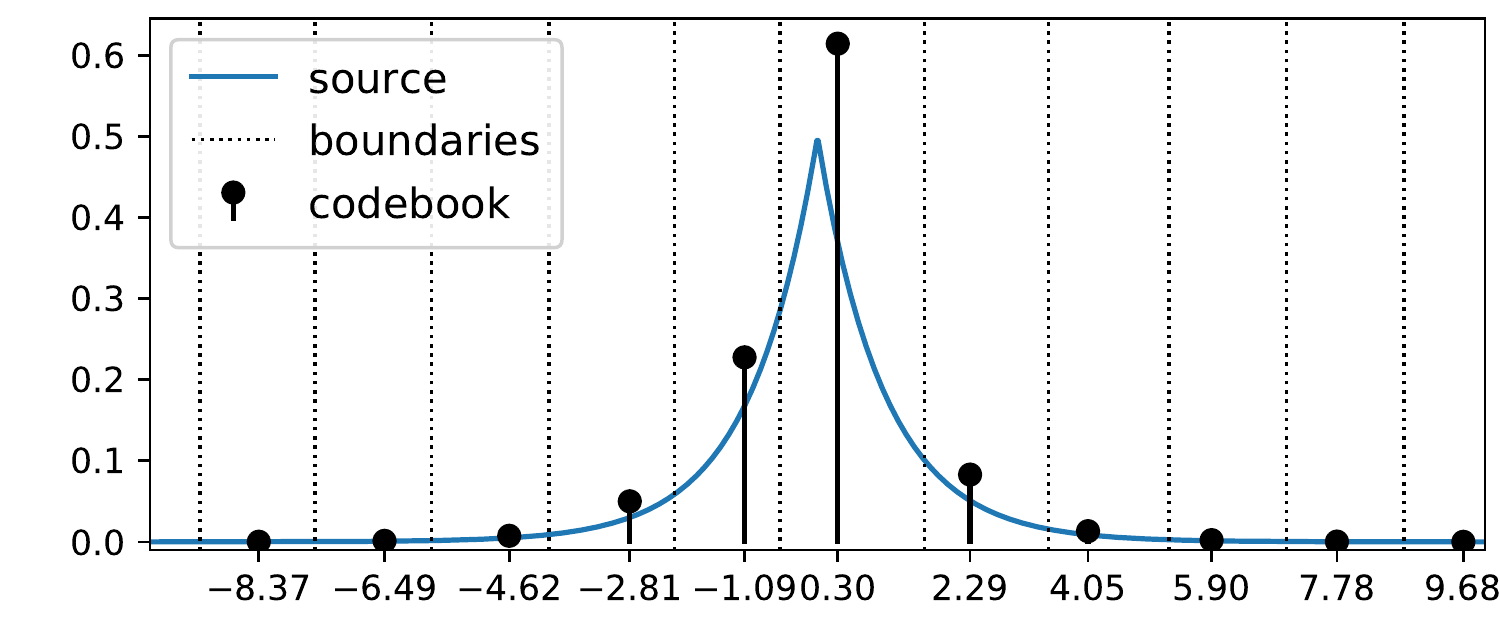}

  \includegraphics[width=\linewidth]{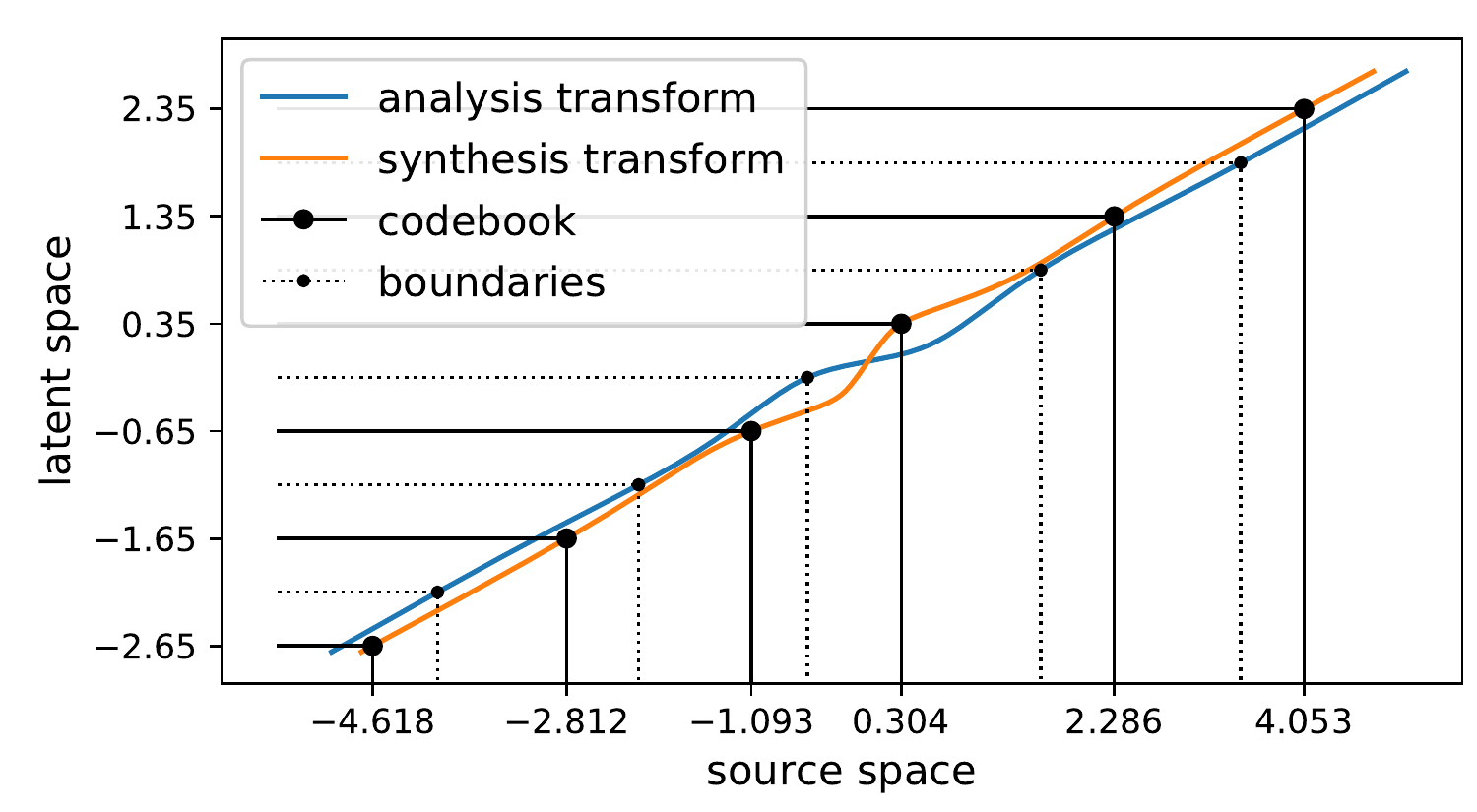}

  \includegraphics[width=\linewidth]{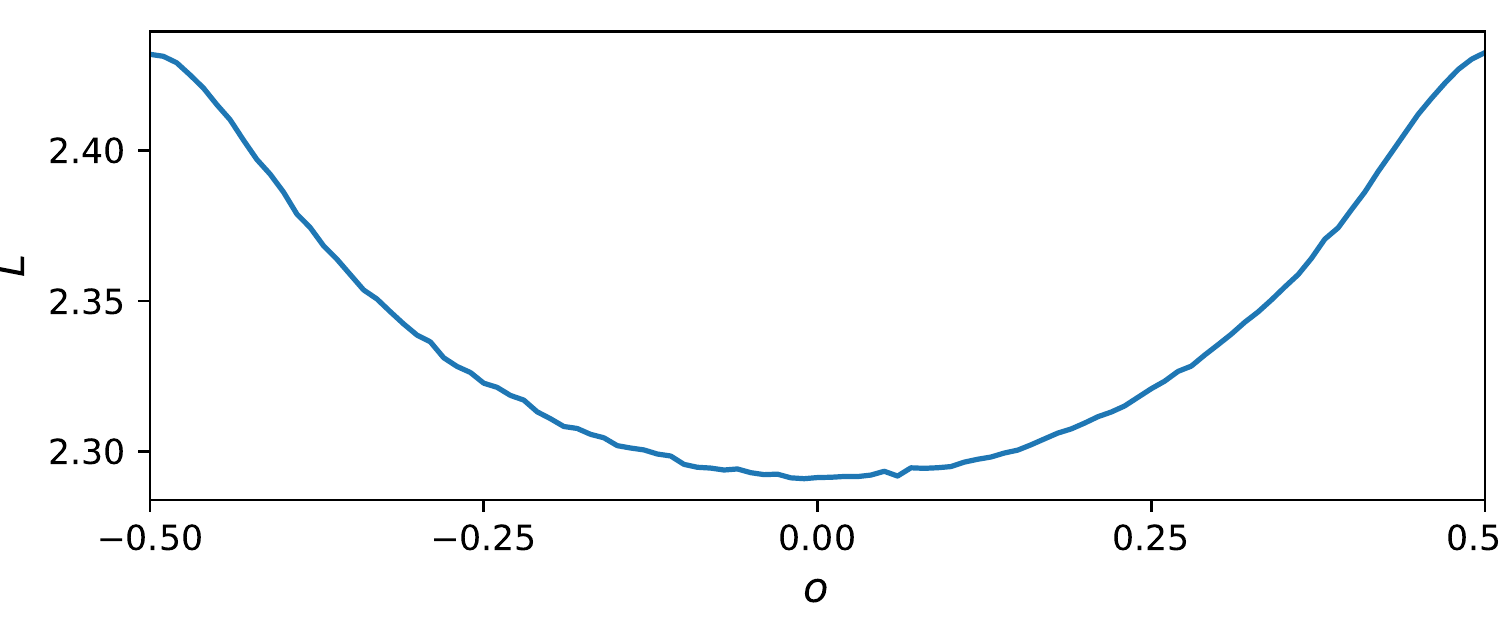}
  \caption{The same instance of NTC as in \cref{fig:ntc_laplacian}, obtained by minimizing $\E_{\bm o} L_{\text{NTC}, \bm o}$. For this figure, a sub-optimal offset was chosen post hoc. Top: visualization of effective quantizer. Center: analysis and synthesis transforms giving rise to the quantizer. Note that the transforms themselves are identical to the ones in \cref{fig:ntc_laplacian}, but the quantization in the latent space is performed with an offset $o = .35$. Bottom: $L_{\text{NTC},o}$ as a function of $o$.}
  \label{fig:ntc_laplacian_offset}
\end{figure}

There is one caveat with using dithered quantization for compression itself: it is not necessarily optimal, since $\E_{\bm o} L_{\text{NTC}, \bm o} \ge \min_{\bm o} L_{\text{NTC}, \bm o}$ (\cref{fig:rd_laplace_banana}, left panel, verifies this empirically). If we do not wish to use dithered quantization, we can still use \cref{eq:loss_noisy} as a proxy loss for transform coding with a fixed quantization offset known to both Alice and Bob.
A simple algorithm for stochastic optimization of an NTC model is:
\begin{enumerate}
\item Minimize \cref{eq:loss_noisy}.
\item Determine which offsets $\bm o$ minimize $L_{\text{NTC}, \bm o}$. If the continuous entropy model $p$ is accurate enough, this can be done without re-estimating the discrete entropy models, since $P(\bm k; \bm o) = p(\bm k + \bm o)$.
\end{enumerate}
Note that the offsets themselves cannot be determined via gradient descent on $L_{\text{NTC}}$, for the same reason that $g_a$ cannot be determined in this way. We must therefore use some other method. While \citet{BaLaSi17} explicitly perform a grid search over $\bm o$, some follow-up papers have resorted to a simple heuristic: guided by the result that for Laplacian distributions, it is optimal to pick an offset that aligns the mode of the source distribution with a codebook vector, one can simply pick an offset for each latent dimension such that one of the quantization bins is centered on the mode (or, in case that is computationally intractable, the median) of the entropy model $p$ \citep{BaMiSiHwJo18,MiBaTo18}. For an entropy model with fixed mode, such as a zero-mean Gaussian, this implies that the offset may as well be fixed a priori.

A suboptimal choice of offset for an NTC encoding a Laplace source is illustrated in \cref{fig:ntc_laplacian_offset}, along with a plot of $L_{\text{NTC}, o}$ as a function of $o$ for the same source. Note that optimizing the dithering proxy loss leads to the transforms becoming increasingly curved around the central quantization bin, to accommodate arbitrary choices of $o$. Because the analysis transform becomes increasingly flat around the center, and the synthesis transform increasingly steep, the effective code vector and quantization bin around the mode of the distribution is skewed towards the near-optimal quantizer illustrated in \cref{fig:ntc_laplacian}. It could be argued that, to minimize the loss function, this should happen in all bins. However, we haven't observed this empirically, presumably due to ANNs naturally favoring smoother functions, and the other bins not contributing enough to the value of the loss function.

\begin{figure}
  \includegraphics[width=\linewidth]{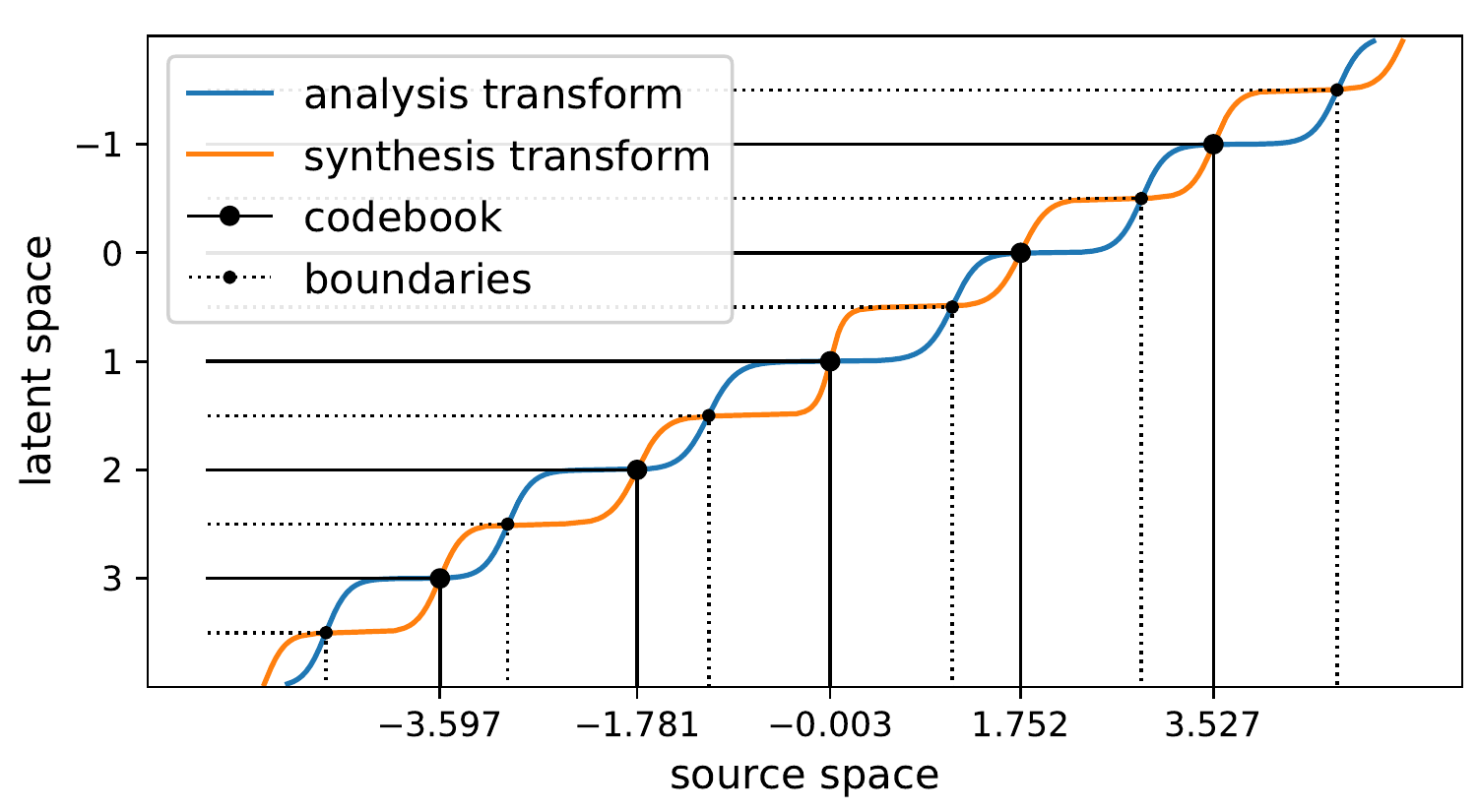}

  \includegraphics[width=\linewidth]{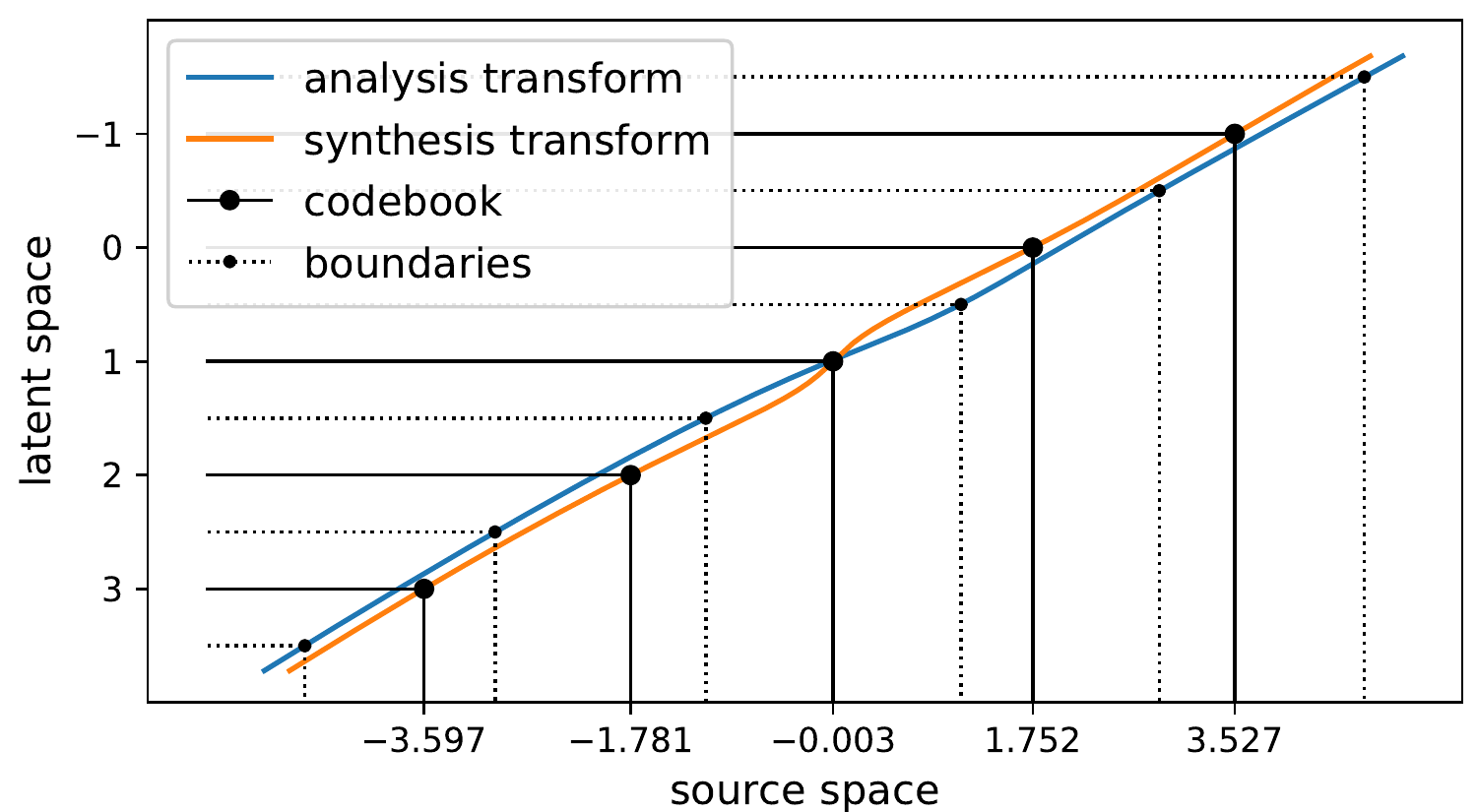}
  \caption{Transforms for Laplace source with explicit soft quantization due to \citet{AgTh20}. Top: ANN transforms including the explicit soft quantization. Bottom: ANN transforms excluding the explicit soft quantization. With this technique, the ANNs themselves can implement smoother (i.e., in some sense, simpler) functions.}
  \label{fig:ntc_laplacian_softquant}
\end{figure}

\citet{AgTh20} discuss augmenting the transforms with a soft quantization function (and making appropriate modifications to the entropy model), which explicitly implements the curvature observed in \cref{fig:ntc_laplacian}. The soft quantization function has a \emph{temperature} parameter, interpolating between the identity function and hard quantization. By explicitly modeling this behavior, the technique relieves the ANN itself from implementing it (\cref{fig:ntc_laplacian_softquant}), and represents a more controlled approach to bridging the gap between quantization and additive uniform noise while retaining near-optimal performance at all rates (\cref{fig:rd_laplace_banana}). The temperature parameter allows explicitly trading off the bias of the proxy loss with the variance of the gradients. The method also suggests that the hard quantization offset after training can simply be chosen to be consistent with the soft quantization offset during training. However, it requires careful choice of an annealing schedule for the temperature. For simplicity, the experiments in this paper use the mode-centering approach described above.

\begin{figure}
  \includegraphics[width=\linewidth]{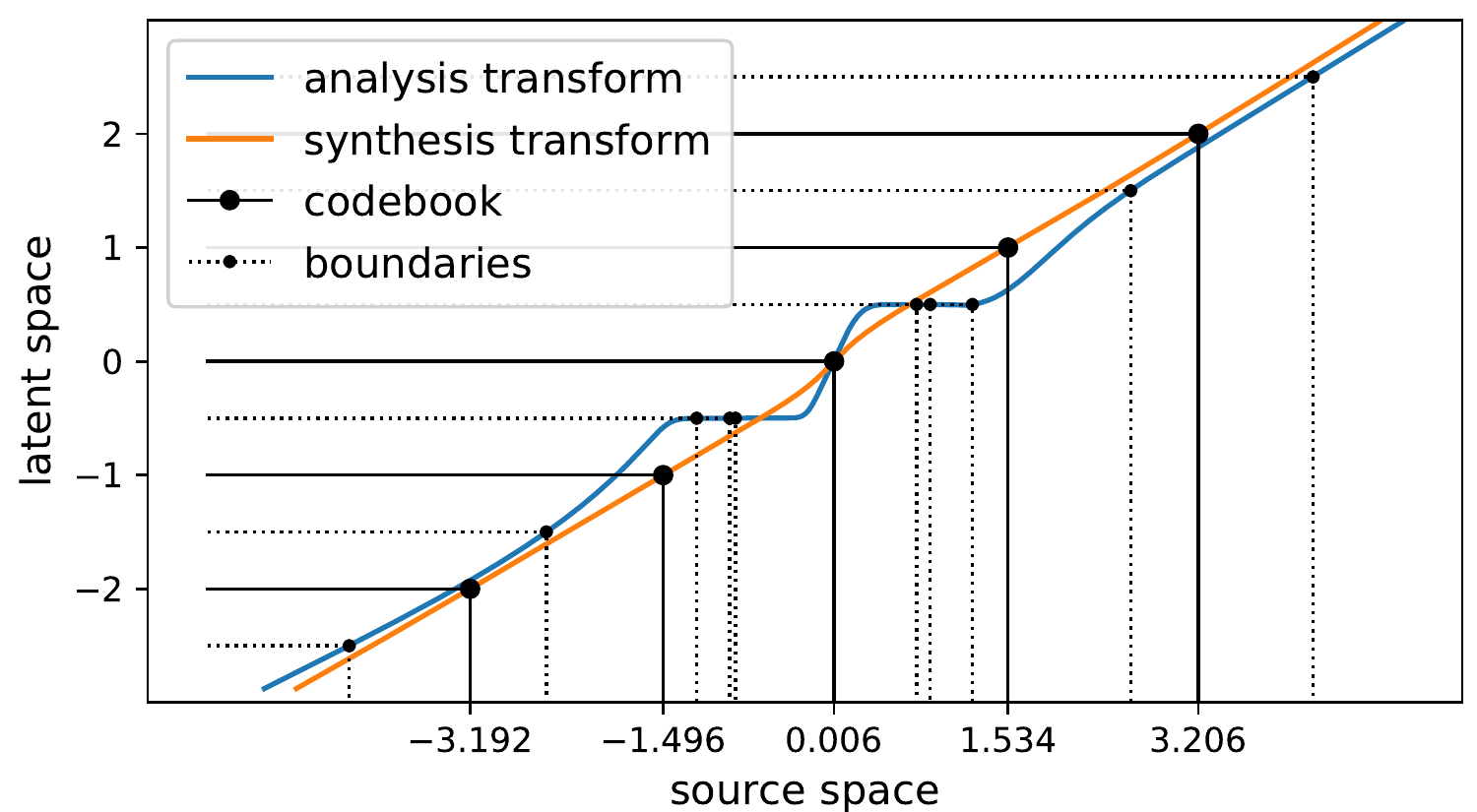}
  
  \includegraphics[width=\linewidth]{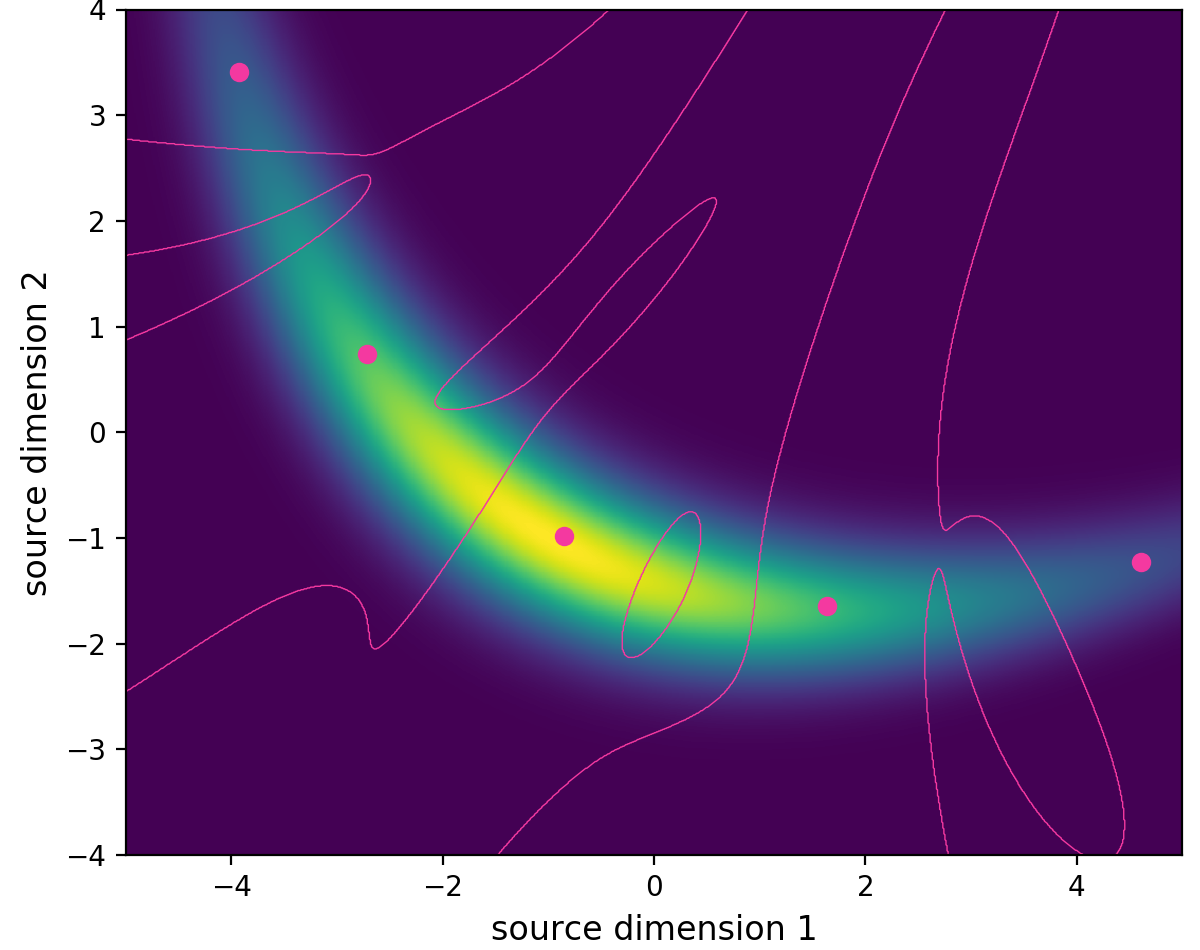}
  \caption{Instabilities observed with straight-through proxy objective at low rates for the Laplace (top) and banana (bottom) distribution. While the synthesis transform tends to be smoother, the analysis transform begins oscillating around the locations of bin boundaries, leading effectively to discontiguous quantization bins.}
  \label{fig:st_instabilities}
\end{figure}

Other authors choose to retain the dithering proxy for the rate term, but use a \emph{straight-through} gradient estimate for the distortion term (effectively computing the distortion loss with constant-offset quantization during training, but replacing the gradient expression of the quantization operation with the identity function \citep{agustsson2020scale,oktay2019scalable,singh2020featcomp,MiSi20}. We have found this approach to yield reasonable results at higher rates, but at low rates, the ad-hoc nature of this approach leads to problematic behaviors of the transforms (RD performance plotted in \cref{fig:rd_laplace_banana}, illustration in \cref{fig:st_instabilities}).

\subsection{Nonlinear transforms}
\label{sec:transforms}
ANNs are known as universal function approximators, and as such we permitted ourselves to ignore their details in the examples above. However, it is crucial to take into account their limitations, some of which arise as a function of their architecture. This is particularly important for complex or high-dimensional source distributions, such as natural images. With higher complexity and rates (larger values of $\lambda$), the optimal transforms generally are more complex and require neural networks with an increasing number of parameters.\footnote{It could be argued that in the high-rate limit, the transforms should collapse to identity functions. However, we haven't observed this effect for image compression models and practically interesting rate--distortion trade-offs, suggesting that this is only the case for extremely high rates.}

In general, neural networks are compositions of layers (parametric functions $\R^A \to \R^B$), wherein each layer typically consists of a linear transformation such as matrix multiplication or convolution, followed by the addition of a bias vector, followed in turn by a nonlinear function, which is typically applied separately on each vector dimension:
\begin{equation}
\bm v = g(\bm r), \text{ with } \bm r = \bm W \bm u + \bm b,
\end{equation}
where $\bm u \in R^A$ is the input vector to the layer, $\bm v \in R^B$ are the layer's outputs or \emph{activations}, and $\bm W \in \R^{B \times A}$ and $\bm b \in \R^B$ are the layer's parameters. For the NTC examples above, we used neural networks with 4 fully connected (i.e., non-convolutional) layers, the first three using the $\softplus$ nonlinearity ($g(x) = \ln(1 + e^x)$ elementwise), while the last layer omits the nonlinearity in order not to constrain the range of the transform to positive values. The \emph{approximation capacity}, i.e., the capability of the neural network to approximate increasingly complex functions, grows with the number of units per layer ($A$, $B$), as well as the depth of the network (the number of layers). Above, we chose $A = B = 100$ (except that we set $A = N$ for the first, and $B = M = N$ for the last layer in $g_a$; analogous for $g_s$), which we found empirically to be large enough for all chosen values of $\lambda$.

For practical sources such as images, video, or audio, imposing special structure in the transforms may have significant benefits in terms of computational complexity, training data efficiency, or both. Generally, NTC models for this type of data use combinations of architectural constraints, most commonly \emph{convolutionality} in $g_a$ and $g_s$, as well as \emph{downsampling} in $g_a$ and \emph{upsampling} in $g_s$, making the transforms share certain characteristics with multi-scale filterbanks, and leading to latent vectors with a tensor structure, consisting of one or more spatial/temporal dimensions, as well as one \emph{channel} dimension (akin to subbands). A detailed example of such an architecture is described by \citet{BaLaSi17}.

\begin{figure}
  \includegraphics[width=\linewidth]{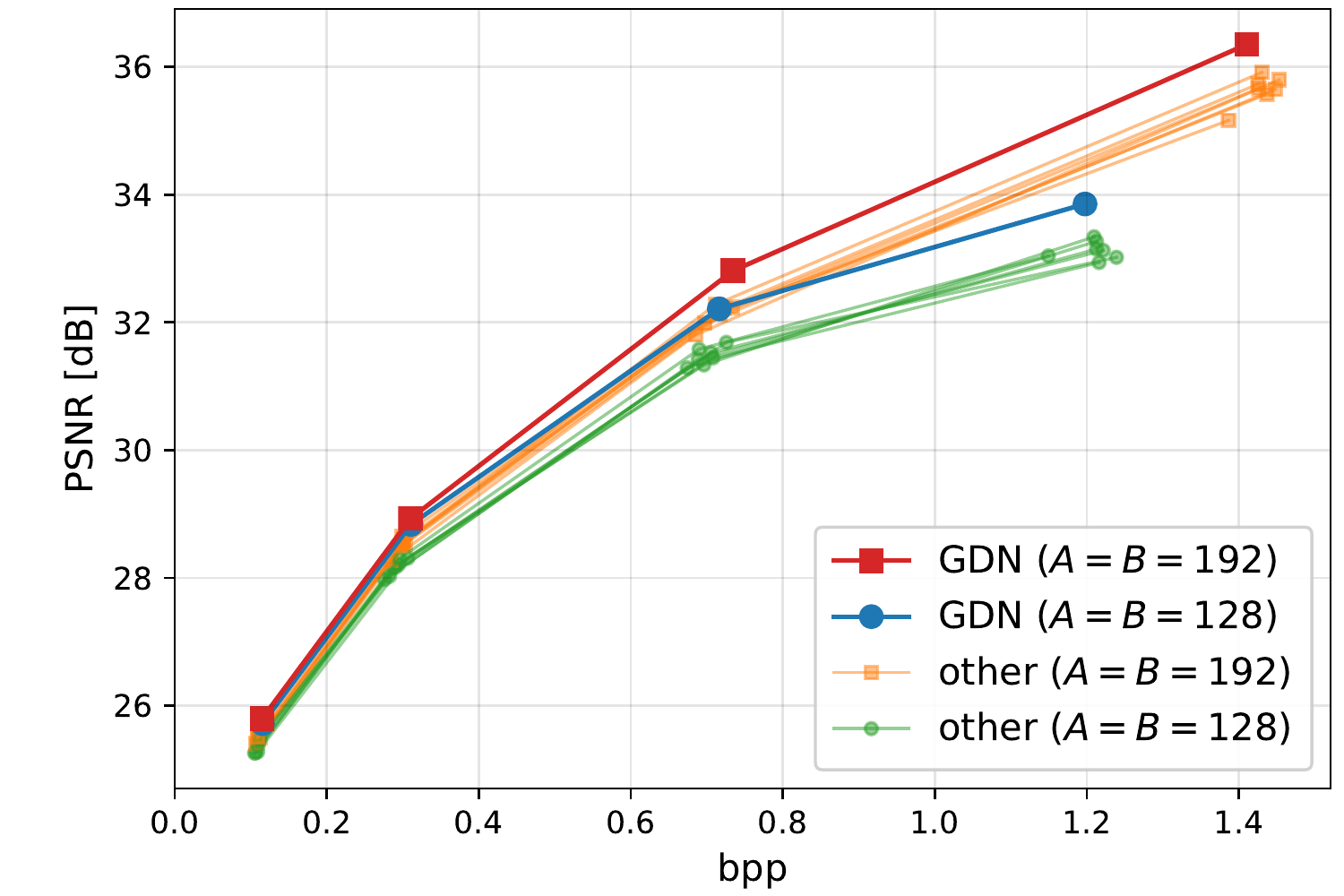}
  \caption{Rate--distortion performance of an NTC model with GDN vs. a collection of pointwise nonlinearities on the Kodak testset \citep{Kodak} (after \citet{Ba18}, with more nonlinearities: ReLU, leaky ReLU, tanh, softplus, ELU \citep{ClUnHo16}, SELU \citep{KlUnMaHo17}, Swish \citep{RaZoLe17}). We compare networks with a different number of hidden units $A, B$ per layer. At low rates, the approximation capacity of all networks is sufficient, and performance converges. At high rates, the capacity of smaller networks saturates earlier; in this regime, the performance benefit of GDN vs. other activation functions becomes measurable.}
  \label{fig:gdn_capacity}
\end{figure}

It has been observed that spatially local normalization as a nonlinearity is beneficial in terms of the trade-off between number of units and RD performance in image compression. In particular, a computationally optimized version of generalized divisive normalization (GDN) \citep{BaLaSi16} as used in recent models is defined as
\begin{equation}
\label{eq:gdn_l1}
v_i = \frac {r_i} {\beta_i + \sum_j \gamma_{ij} \, |r_j|},
\end{equation}
where $\bm r$ are the linear responses of the layer, $\bm v$ represents the vector of normalized responses (the activations), and the vector $\bm \beta$ and matrix $\bm \gamma$ represent parameters of the transformation (both non-negative). The computation is typically replicated across spatial dimensions, as linear filtering is in convolutions, and $i, j$ only index the channel dimension. \citet{JoEbGoBa19} show that the originally more complex form of GDN can be simplified to resemble a weighted $\ell^1$-norm (plus a constant), as in \cref{eq:gdn_l1}, with negligible RD performance loss, but minimizing computationally costly exponentiations. Since ANNs can be understood as universal function approximators, the benefit of a particular architectural constraint may only become evident when the network is at its approximation capacity. \citet{Ba18} finds that this is the case at higher rates; i.e., for a constant network architecture, the superiority of GDN vs. pointwise nonlinearities disappears at lower rates (\cref{fig:gdn_capacity}). \citet{JoEbGoBa19} also discuss the trade-off between computational complexity and RD performance resulting from other architectural choices in further detail, such as the number of channels per layer or the number of decoder layers, and provide an algorithm to semi-automatically determine these hyperparameters.

\section{Learned entropy models}
\label{sec:entropy_modeling}
In linear transform coding with a Gaussian source assumption, the probabilistic model $P$ in \cref{eq:loss_ntc} is typically considered to be a distribution factorized over each latent dimension, since the KLT factorizes the source.
However, as pointed out by \citet{Go01}, this is not necessarily a good model for real-world sources, and decorrelating the source is not generally RD optimal; this is also illustrated in \cref{fig:rd_laplace_banana}.

In NTC, any differentiable density model $p$ can be used to minimize \cref{eq:loss_noisy} in principle; after determining a quantization offset, it can be translated into a corresponding probability mass function $P(\cdot; \bm o)$, and used for entropy coding. However, to fully benefit from the function approximation capabilities of ANNs, techniques have been developed that allow jointly optimizing the transforms with an ANN-based density model, such that both components of the model are adapted to each other as best as possible. Key to this approach is conditioning. Generally, entropy coding methods such as arithmetic coding process one dimension at a time. Thus, we must be able to write the entropy model as a chain of conditionals:
\begin{equation}
P(\bm{\hat y} \mid \bm{\hat z}) = \prod_i P(\hat y_i \mid \bm{\hat y}_{:i}, \bm{\hat z}),
\end{equation}
where $\bm{\hat y}$ is the quantized latent representation, $\bm{\hat y}_{:i}$ indicates the vector comprising the dimensions of $\bm{\hat y}$ preceding the $i$th (according to some predetermined ordering), and $\bm{\hat z}$ is another (optional) vector that must be known to both Alice and Bob. In the simplest case, $P$ is assumed factorized, and the chain collapses to a product of independent scalar densities. Conditioning on some other vector $\bm{\hat z}$ typically requires transmitting the vector as side information, and thus corresponds to forward adaptation (FA) of the density model. Conditioning on the preceding dimensions of $\bm{\hat y}$ can be done without additional side information, but requires interleaving the computation of the probabilities with the decoding of $\bm{\hat y}$; this is backward adaptation (BA).
FA and BA have long been used in conventional image compression.  Context-adaptive arithmetic coding is an example of BA; mode selection and signaling is an example of FA \citep{MaScWi03}.

\begin{figure}
  \centering
  \begin{tikzpicture}[x=.9em,y=-1.2em]
    \node[image] at (0,0) (x) {$\bm x$};
    \node[image] at (0,18.5) (x_hat) {$\bm{\hat x}$};

    \node[nn, shape border rotate=270, minimum size=2em] at (4.5,0) (g_a) {$g_a$};
    \node[nn, shape border rotate=270, minimum size=2em] at (4.5,18.5) (g_s) {$g_s$};

    \node[tensor] at (9,0) (y) {$\bm y$};
    \node[tensor] at (9,18.5) (y_hat) {$\bm{\hat y}$};

    \node[nn, shape border rotate=270, minimum size=2em] at (16,0) (h_a) {$h_a$};
    \node[nn, shape border rotate=180, minimum size=1em] at (14,16) (h_m) {$h_m$};
    \node[nn, shape border rotate=180, minimum size=1em] at (18,16) (h_s) {$h_s$};

    \node[tensor] at (22,0) (z) {$\bm z$};
    \node[tensor] at (22,18) (z_hat) {$\bm{\hat z}$};

    \node[tensor] at (14,13) (mu) {$\bm \mu$};
    \node[tensor] at (18,13) (sigma) {$\bm \sigma$};

    \node[op] at (9,3) (y_om) {$+$};
    \node[op, minimum size=3ex] at (9,5) (y_q) {\footnotesize $\lfloor \cdot \rceil$};
    \node[filter] at (9,7) (y_ae) {EC};
    \node[filter] at (9,11) (y_ad) {ED};
    \node[op] at (9,13) (y_op) {$+$};

    \node[op] at (22,3) (z_om) {$+$};
    \node[op, minimum size=3ex] at (22,5) (z_q) {\footnotesize $\lfloor \cdot \rceil$};
    \node[filter] at (22,7) (z_ae) {EC};
    \node[filter] at (22,11) (z_ad) {ED};
    \node[op] at (22,13) (z_op) {$+$};
    \node[input, right] at (24,3) (om) {$\bm o$};
    \node[input, right] at (24,13) (op) {$\bm o$};

    \draw[->, thick] (x) -- (g_a) -- (y);
    \draw[->, thick] (y_hat) -- (g_s) -- (x_hat);

    \draw[->, thick] (y) -- (h_a) -- (z);
    \draw[->, thick] (z_hat) -| (h_s.290);
    \draw[->, thick] (z_hat-|h_s.290) node[branch] {} -| (h_m.290);
    \draw[->, thick] (h_m) -- (mu) -- (y_op);
    \draw[->, thick] (mu) |- (y_om) node[below right=.3ex] {\footnotesize $-$};
    \draw[->, thick] (h_s) -- (sigma) |- (y_ad);
    \draw[->, thick] (sigma|-y_ad) node[branch] {} |- (y_ae);

    \draw[->, thick] (y) -- (y_om) -- (y_q) -- (y_ae) -- (y_ad);
    \draw[->, thick] (y_ad) -- (y_op) -- (y_hat);
    \draw[->, thick] (z) -- (z_om) -- (z_q) -- (z_ae) -- (z_ad);
    \draw[->, thick] (z_ad) -- (z_op) -- (z_hat);
    \draw[->, thick] (om) -- (z_om) node[below right=.3ex] {\footnotesize $-$};
    \draw[->, thick] (op) -- (z_op);

    \draw[->, thick, densely dotted] (y_hat) -| (h_m.250) node[below right,pos=.1] {$\bm{\hat y}_{:i}$};
    \draw[->, thick, densely dotted] (y_hat-|h_m.250) node[branch] {} -| (h_s.250);

    \node[right] at (-2,9) (channel) {channel};
    \draw[thick, dotted] (channel) -- (26,9);

    \begin{pgfonlayer}{background}
      \node[fill=ggreen, opacity=.2, rounded corners=2ex, fit=(y_hat)(z_hat)(y_ad)(z_ad)(op.mid), inner xsep=1.5ex, inner ysep=1.5ex] {};
    \end{pgfonlayer}
  \end{tikzpicture}
  \caption{Illustration of a nonlinear transform coder using both learned forward and backward adaptation. $\bm x$ is the source vector, $\bm{\hat x}$ the reconstruction. $\bm y$ is a latent representation tensor, and $\bm{\hat y}$ its uniformly quantized counterpart. $\bm z$ and $\bm{\hat z}$ are an analogous hierarchical latent representation computed via a transform $h_a$, representing side information. While the entropy model on $\bm{\hat z}$ is predetermined, the entropy model on $\bm{\hat y}$ is here assumed conditionally independent Gaussian with mean tensor $\bm \mu$ and standard deviation tensor $\bm \sigma$. Both tensors are computed as functions of previously decoded values of $\bm{\hat y}$ (backward adaptation) and the side information $\bm{\hat z}$ (forward adaptation) using the ANNs $h_m$ and $h_s$. While Bob begins with entropy decoding $\bm{\hat z}$ (ED), and then uses it to decode $\bm{\hat y}$, Alice must have access to the entropy model on $\bm{\hat y}$ to entropy encode it (EC). Thus, in addition to computing the upper half of the diagram, she must also compute the section in the shaded box. The quantization offset of $\bm y$ is assumed aligned with the conditional mean of the Gaussian, making the entropy model only dependent on $\bm \sigma$. To enable reliable cross-platform decoding of $\bm{\hat y}$, $h_s$ may be computed using a learned integer transform and translated into an arithmetic code via lookup tables.}
  \label{fig:system_diagram}
\end{figure}
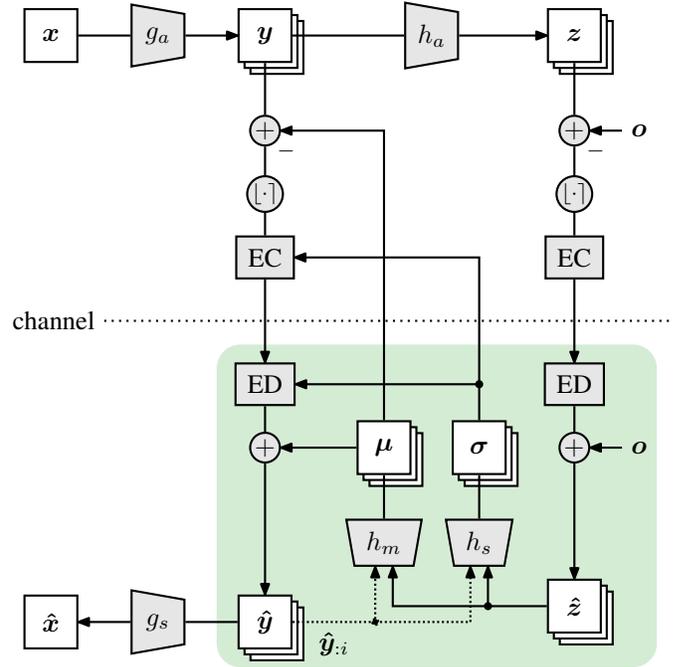

\begin{figure*}
  \includegraphics[width=.49\linewidth]{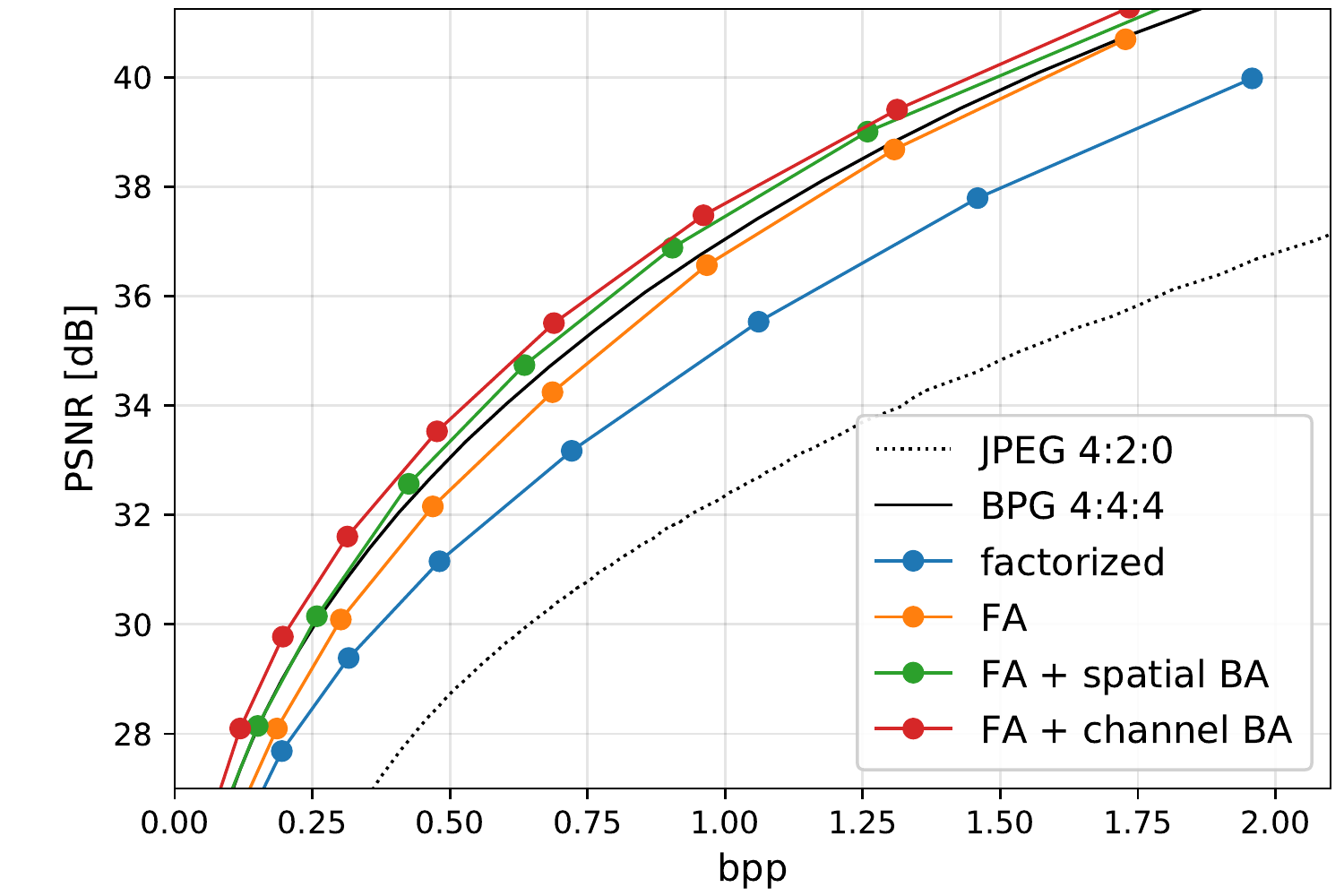}\hfill%
  \includegraphics[width=.49\linewidth]{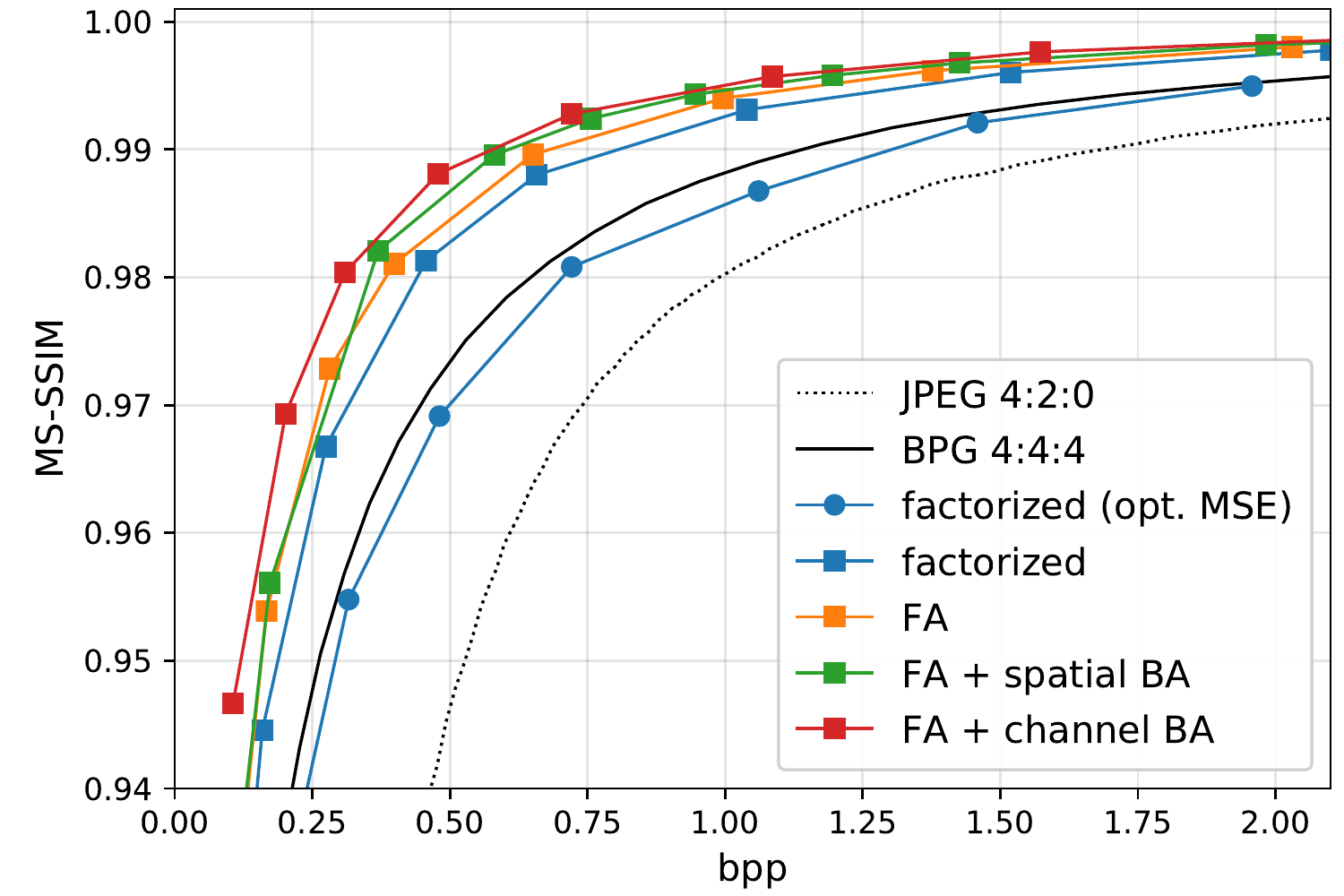}
  \caption{Image compression performance of NTC on the Kodak testset \citep{Kodak}, comparing different learned entropy models to JPEG and BPG, which is a popular variant of HEVC, a relatively recent and popular commercial method. The \emph{factorized} and \emph{FA} models replicate \citet{BaMiSiHwJo18}; the \emph{FA + spatial BA} model is due to \citet{MiBaTo18}; the \emph{FA + channel BA} model is provided by \citet{MiSi20}. We compare models that are approximately equivalent in terms of optimization procedure and architecture of analysis and synthesis transforms. Conditional entropy models lead to significant improvements over factorized models. With sophisticated entropy modeling, learned image compression compares favorably to BPG in terms of PSNR, which BPG is optimized for. Regarding MS-SSIM, note that even the factorized model optimized for mean squared error (MSE) performs relatively closely to BPG, and when optimized for MS-SSIM, far outperforms it. We attribute this to the fact that regardless of the distortion measure, nonlinear transforms are better suited to model the source distribution (\cref{fig:ntc_banana}), and that MS-SSIM captures certain characteristics of the source that are also relevant perceptually (\cref{fig:kodim15}).}
  \label{fig:entropy_modeling}
\end{figure*}

Learned forward adaptation was first described in the context of image compression \citep{BaMiSiHwJo18}, inspired by the observation that the magnitudes of spatially nearby elements of the latent tensor in a convolutional NTC for images tend to be correlated. As illustrated in \cref{fig:system_diagram}, $\bm y = g_a(\bm x)$ is further processed by an ANN $h_a$ to produce a side information vector $\bm z$. The entropy model for $\bm{\hat y}$ is conditioned on the decoded $\bm{\hat z}$. \citet{BaMiSiHwJo18} assume elements of $\bm y$ to be zero-mean Gaussians, conditionally independent wrt. $\bm{\hat z}$; a non-zero mean model is introduced by \citet{MiBaTo18}. The entropy of $\bm{\hat z}$ is small enough to warrant the improved fit of $P(\bm{\hat y} \mid \bm{\hat z})$, effectively lowering the rate. \Cref{fig:entropy_modeling} compares the rate--distortion performance of several learned image compression models with JPEG \citep{JPEG} and BPG, a variant of HEVC \citep{HEVC}, in terms of peak signal-to-noise ratio (PSNR) as well as MS-SSIM, a popular perceptual image quality metric \citep{WaSiBo03}. The introduction of FA into NTC leads to a significant improvement of RD performance with respect to both metrics.

\citet{MiBaTo18} are among the first authors to discuss learned backward adaptation. They introduce a spatially autoregressive model, where $\bm{\hat y}$ is processed one spatial location at a time, producing a distribution for each channel vector conditioned on previously decoded spatial locations. Combined with FA, the model produces further RD gains over the FA-only model (\cref{fig:entropy_modeling}), and outperforms BPG.

A downside of BA compared to FA is that it impedes computational parallelism: the system must alternate between computing conditional probabilities and entropy decoding. On the other hand, with FA, Alice may effectively convey more information to Bob than necessary.\footnote{We can write the excessive information as $H(\bm{\hat z} \mid \bm{\hat y})$, i.e. the conditional entropy of the side information $\bm{\hat z}$ given the latents $\bm{\hat y}$. It is also referred to as \emph{bits-back} cost \citep{Wa90,HiZe93}; practical algorithms to asymptotically get these “bits back” have been proposed by \citet{FrHi97} and more recently by \citet{ToBiBa19}. Note that in all models compared in \cref{fig:entropy_modeling}, the side information only amounts to a fraction of the total bit rate. Thus, since $H(\bm{\hat z}) \ge H(\bm{\hat z} \mid \bm{\hat y})$, the bits-back cost in these models is negligible.} To alleviate the computational bottleneck in BA, \citet{MiSi20} introduce a model that iterates over channel slices rather than spatial locations, which is more amenable to parallelization using GPUs and, along with further modeling improvements, presents a significant improvement over traditional methods.

A problem frequently encountered with conditional entropy models is numerical determinism. To make image compression models practically relevant, they need to be implemented on a wide variety of hardware platforms. However, when probabilities are computed using floating-point arithmetic, numerical round-off errors can lead to catastrophic decoding failures due to the sensitivity of entropy coding with respect to discrepancies in the probability model between sender and receiver. The exact numerical round-off at each layer of an ANN depends on the hardware representation of floating point numbers, as well as the mode of parallelism, because round-off errors are not associative. This problem is typically handled in linear transform coders by using lookup tables to model probabilities \citep[e.g., ][]{MaScWi03}. \citet{BaMiJo18} provide a solution for ANN-based entropy modeling, where ANNs are trained using floating-point arithmetic, but use integer arithmetic when deployed. This enables reliable decoding on arbitrary hardware platforms for the above-mentioned class of entropy models.

\section{RD traversal with $\lambda$-parameterization}
\label{sec:multirate}
The loss function in \cref{eq:loss_noisy} is optimized in expectation over the source distribution. The resulting transform thus jointly minimizes the rate and the expected distortion $d$ between the source and the reconstruction, for a fixed trade-off predetermined by the choice of $\lambda$. In many linear transform coders, the system is parameterized by the quantization step size, such that only one set of transforms is needed to continuously traverse a range of RD trade-offs. Using this approach with a single set of trained nonlinear transforms was first explored by \citet{dumas2018autoencoder}. However, this is not generally optimal.

With NTC, the transforms and/or the entropy model can be made more general functions of $\lambda$. One method to “condition” an ANN, first introduced in the context of stylization tasks \citep{DuShKu17}, is to insert additional computations between layers, such as affine transformations:
\begin{equation}
\bm w = h_f(\lambda) \odot \bm v + h_b(\lambda),
\label{eq:lambda_parameterization}
\end{equation}
where $\bm v$ are the outputs of a layer, $\bm w$ are the inputs to the next layer, and $\odot$ represents elementwise multiplication. In this context, $h_f$ and $h_b$ are parametric functions of $\lambda$ that can be computed themselves via ANNs. The parameters of these ANNs in turn are optimized for the RD objective (\cref{eq:loss_noisy}) as well. Parameterizing the entropy model and transforms this way was proposed earlier by \citet{choi2019variable,dosovitskiy2019you}. We simplify this approach here by noting that, since $h_f$ and $h_b$ are functions of a scalar, they may be conveniently defined via first-order splines (i.e., piecewise linear functions). Additionally, we propose to remove the $\lambda$-parameterization of the entropy model, and for transforms using GDN, to simply treat its parameters ($\bm \beta$ and $\bm \gamma$) as functions of $\lambda$ instead of using affine transformations.

\begin{figure*}
  \includegraphics[width=.49\linewidth]{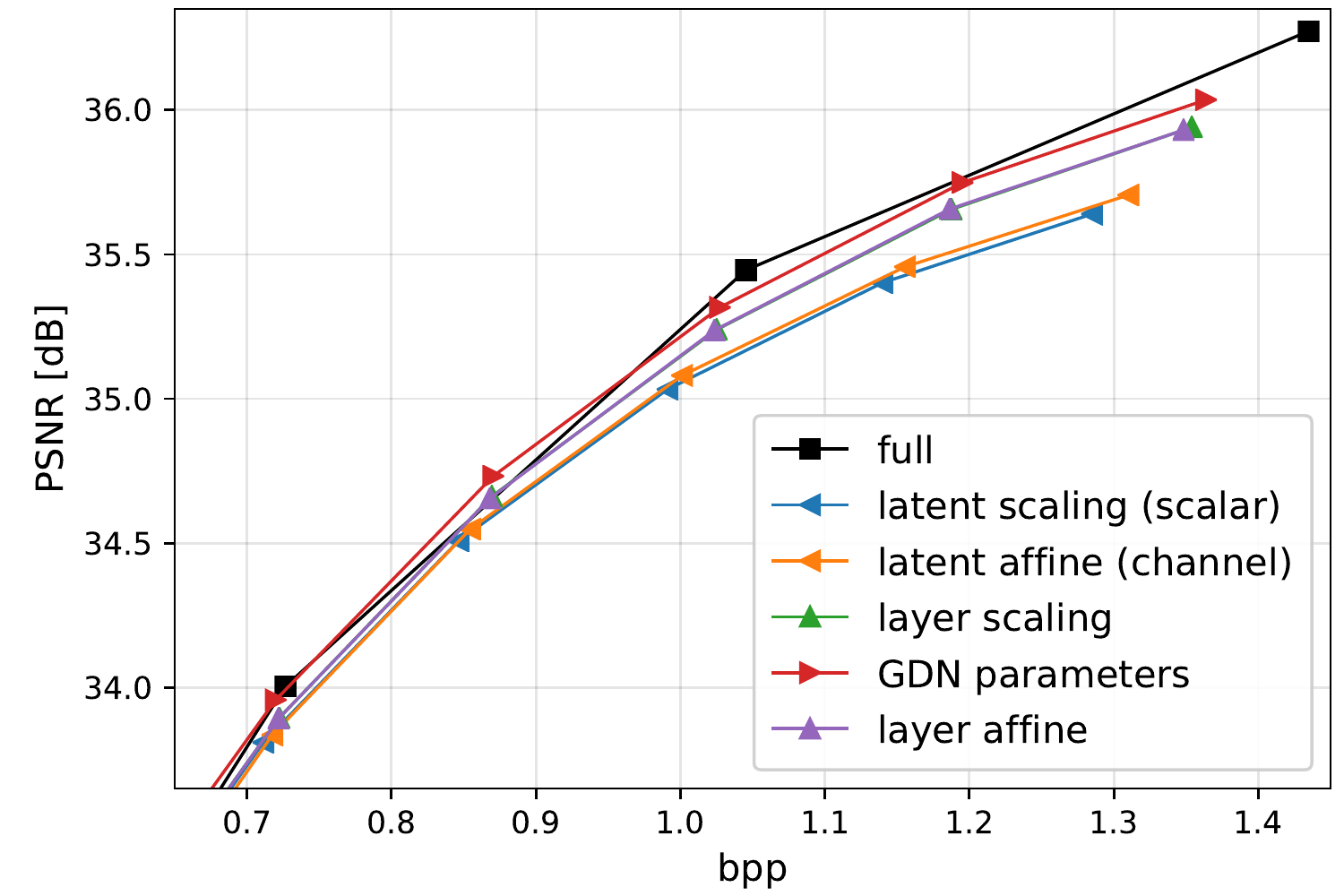}\hfill%
  \includegraphics[width=.49\linewidth]{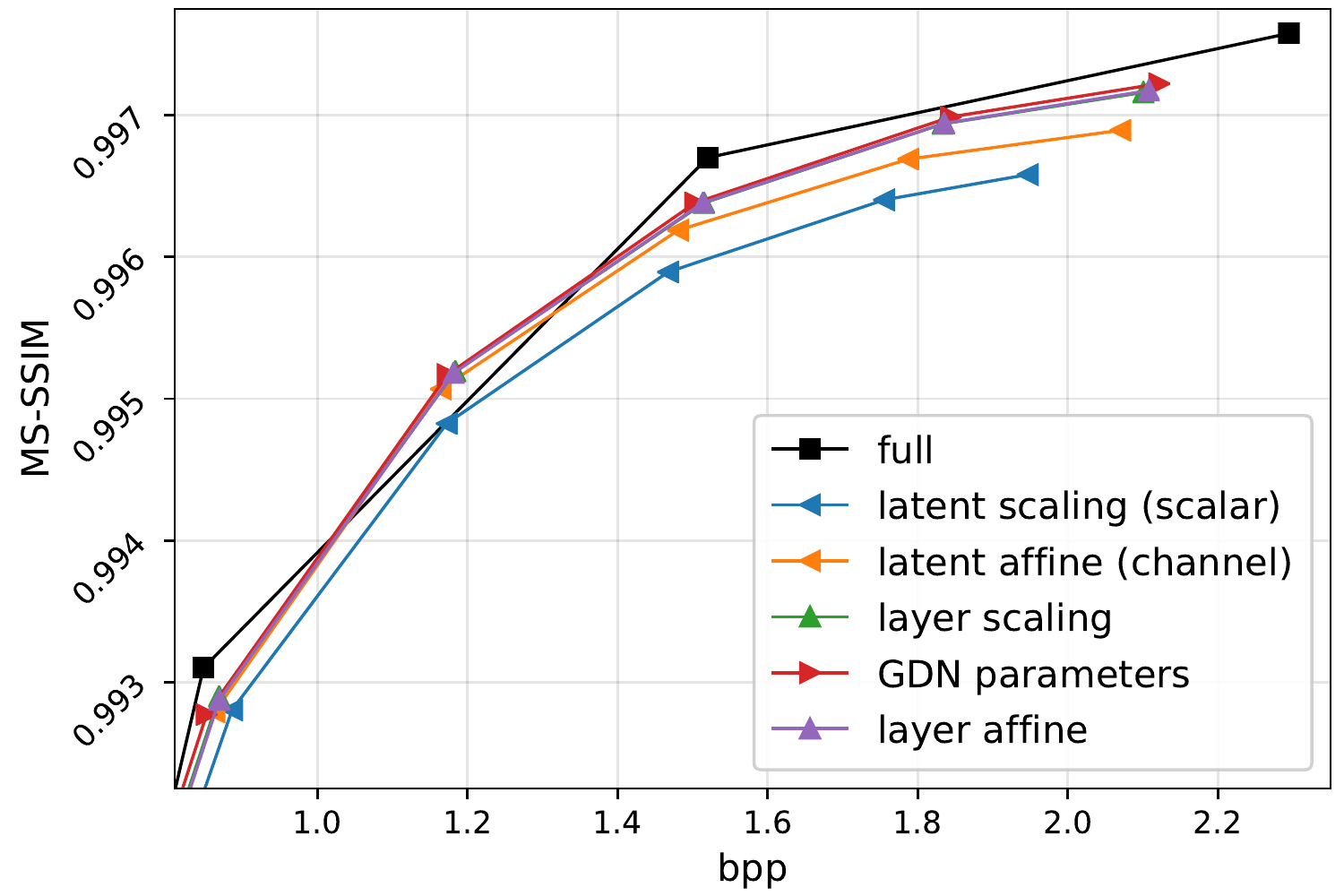}
  \caption{Performance comparison of $\lambda$-parameterization techniques implementing \cref{eq:lambda_parameterization} using first-order splines, and parameterizing only the transforms, not the entropy model. All models have the same architecture and were optimized using the dithering proxy either for MSE (left panel) or MS-SSIM (right panel). Results are shown for the Kodak testset \citep{Kodak}. As in \cref{fig:gdn_capacity}, differences between models become more evident at high rates. Optimization using the straight-through proxy gives consistent results (not shown). \emph{Full}: RD performance of separate models for each $\lambda$; \emph{latent scaling}: only the output of the $g_a$ and the input to $g_s$ are scaled with a single scalar each ($h_b$ is zero); \emph{latent affine}: each output channel of $g_a$ and each input channel of $g_s$ are subjected to a scalar affine transformation, as in \cref{eq:lambda_parameterization}; \emph{layer affine}: each channel of each layer of $g_a$ and $g_s$ is subjected to a scalar affine transformation; \emph{layer scaling}: ditto, except that $h_b$ is zero; \emph{GDN parameters}: rather than adding a scaling between layers, the parameters of each instance of GDN in the transforms ($\bm \beta$ and $\bm \gamma$) are represented as first-order splines dependent on $\lambda$. We note that a simple affine transformation of the latent space, corresponding to varying the quantization interval, is not sufficient to maintain comparable performance with the full model. Scaling the activations of each layer appears sufficient, while reparameterizing GDN as a function of $\lambda$ yields slightly better performance.}
  \label{fig:multirate}
\end{figure*}

\begin{figure}
  \includegraphics[width=\linewidth]{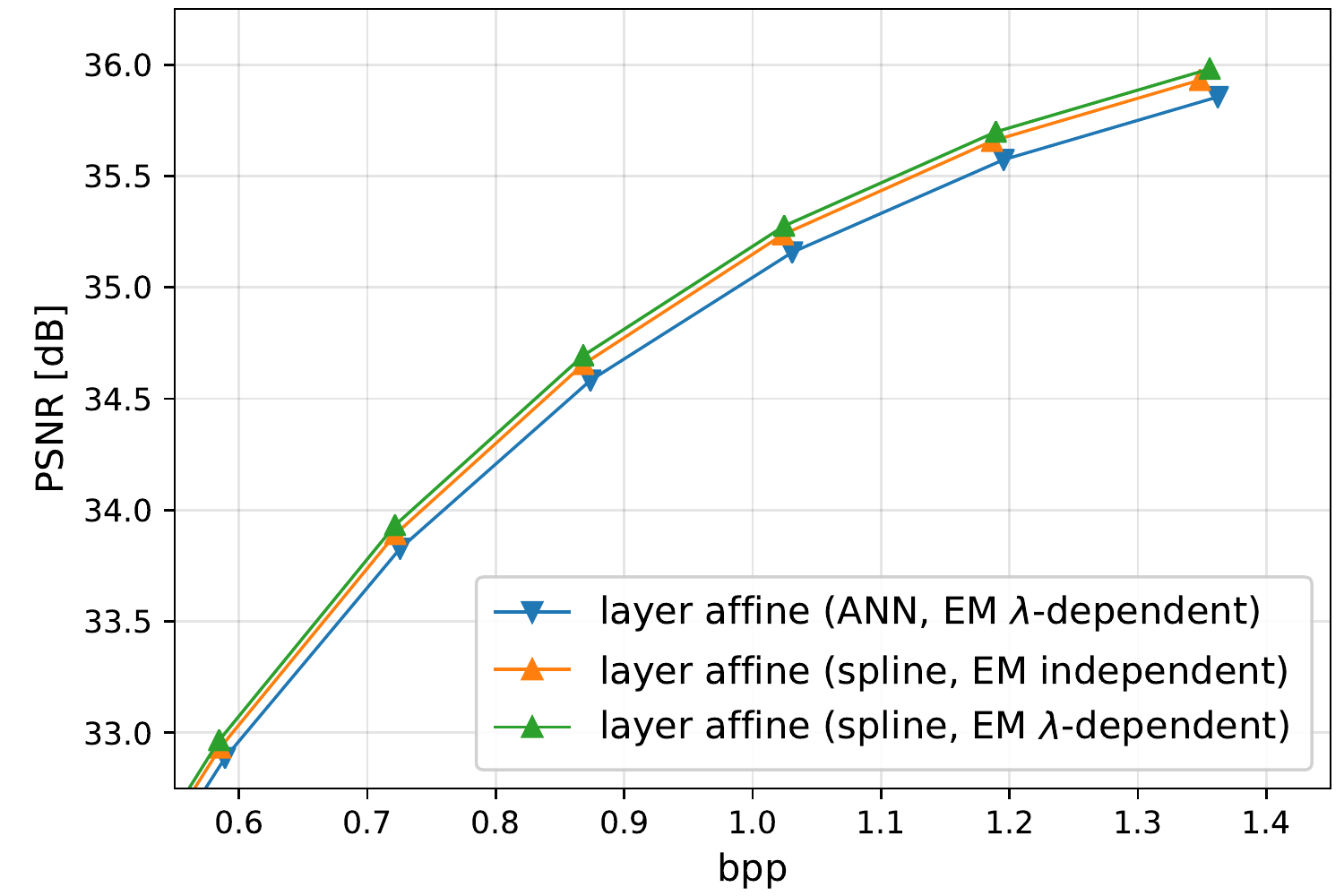}%
  \caption{Performance comparison of $\lambda$-parameterization using splines vs. ANNs, and of optionally applying it to the entropy model. As in \cref{fig:gdn_capacity}, differences between models become more evident at high rates. Comparing models implementing \cref{eq:lambda_parameterization} using ANNs vs. first-order splines, we find that the performance of ANNs is consistently worse than that of splines, despite a larger number of parameters, suggesting that optimization of ANNs may be numerically more difficult in this context. Comparing models using splines and either a $\lambda$-parameterized entropy model, or a forward-adaptive entropy model independent of $\lambda$, we find that the benefit of parameterizing the EM is rather small. Results shown are for the Kodak testset \citep{Kodak} and are optimized using the dithering proxy; the straight-through proxy yields consistent results.}
  \label{fig:multirate_control}
\end{figure}

We carried out experiments with an NTC model with FA following the experimental setup of \citet{BaMiSiHwJo18}, but with 160 channels per layer. When implementing $h_f$ and $h_b$ with ANNs, we used two-layer networks with 128 hidden units for each scalar element produced by the functions. For the spline implementation, we used a first-order spline with 25 parameters. First, we found that the optimization of ANN-based parameterization is numerically more difficult than first-order splines (\cref{fig:multirate_control}). Furthermore, we removed the $\lambda$-parameterization of the entropy model and noted that it is not crucial to RD performance (same figure). Along with the practical requirement that $h_s$ needs to be implemented with integer arithmetic for cross-platform stability, this suggests that making the entropy model explicitly dependent on $\lambda$ may not be worth the complexity of implementation.

\Cref{fig:multirate} compares $\lambda$-parameterizations of only the transforms $g_a$ and $g_s$, and using splines, in terms of RD performance. As for the experiments with different nonlinearities, differences between the parameterizations emerge at high rates, since the network capacities saturate in that regime. We find that a scaling or affine transformation of the latent space alone, roughly equivalent to parameterizing the quantization step size and offset, are not sufficient to achieve an RD performance equivalent to the family of full, non-parameterized models. However, any layer-wise parameterization appears close enough. This is explainable by the fact that the RD family of optimal entropy-constrained scalar quantizers cannot in general be parameterized by a scaling of the quantization offset (a notable exception being the Laplace source discussed above). GDN reparameterization performs the best in an RD sense, but also requires slightly more model parameters compared to the other methods, since \cref{eq:lambda_parameterization} requires two length-$B$ vectors, but $\bm \gamma$ is a $B \times B$ matrix.

\section{Related work}
Due to the resurgence of ANNs and data-driven computing in recent years, the field of data compression has received an influx of new ideas. While transform coding as a concept has been around for decades \citep{AhRa75}, one could observe a recent convergence of it with the idea of autoencoders \citep{HiSa06}. Autoencoders, likewise, have been discussed for decades, but largely in a separate community. One notable step towards this convergence was the fusion of variational Bayesian methods with autoencoders, which introduced a probabilistic interpretation, making the connection to information-theoretic quantities such as entropy \citep{KiWe14,ReMoWi14}; another was the use of a dithering-based loss for optimization of nonlinear transform codes \citep{BaLaSi16a}.

As is often the case in lossy compression, the field of nonlinear transform coding has been driven forward by the need to compress digital images. Early image compression models using ANNs include the work of \citet{ToViJoHwMi17}, who do not use entropy modeling; \citet{RiBo17}, who use a context-based adaptive entropy coder not jointly optimized with the transforms; and \citet{BaLaSi17,ThShCuHu17}, who jointly optimize the transforms with continuous entropy models, the latter with a different formulation than what we use here. \citet{AgMeTsCaTi17} combine an autoencoder with VQ in the latent space over small blocks of coefficients, utilizing a soft quantization proxy. More recent work using soft notions of quantization includes the work of \citet{alexandre2019learned,AgTh20}.

Beyond the use of convolutional filtering, up- or downsampling, and special nonlinearities \citep{BaLaSi17} as discussed earlier, many authors exploit properties of the image distribution by way of introducing special structure into the transforms, such as multi-scale architectures \citep{RiBo17,nakanishi2018neural,cai2018efficient}; non-local, or “attention”-based network architectures \citep{LiChShMa19,ChSuTaKa20a}; or iteration built into the transforms \citep{cai2018deep,ororbia2018learned}. Recently, the topic of extending nonlinear transform codes to video signals has received much attention, and the space of possible network architectures suitable for this application has been explored, including spatiotemporal convolutions, optical flow networks, as well as multi-scale linear filtering \citep{wu2018video,han2018deep,lu2019dvc,chen2019learning,rippel2019learned,habibian2019video,djelouah2019neural,lombardo2019deep,golinski2020feedback,yang2020learning,lu2020end,agustsson2020scale}. \citet{BaMiJo18} develop integer architectures for learned entropy models, in order to guarantee reliable decoding on arbitrary hardware platforms, and \citet{JoEbGoBa19} discuss selecting architecture parameters, such as the number of layers, or number of channels per layer, while taking into account the RD performance.

Notable work in the space of learned entropy models includes \citet{minnen2018image}, which use block-based forward adaptation, and several other concurrent publications performing learned backward adaptation \citep{MiBaTo18,lee2018context,mentzer2018conditional}. More recent work on learned backward adaptation includes the papers by \citet{li2020efficient,guo20203}.

The first work exploring the RD trade-off with a single set of nonlinear transforms is due to \citet{dumas2018autoencoder}, who explore varying the quantization step size, as in linear TC (corresponding most closely to “latent affine” in \cref{fig:multirate}). \citet{guarda2020deep} use the same method for coding point-cloud geometries. Layer-wise parameterization as in \cref{eq:lambda_parameterization} was introduced for image compression by \citet{choi2019variable}. It was generalized by \citet{dosovitskiy2019you} for a range of other tasks including compression. Our technical contribution is to simplify the parameterization and to adapt it for GDN.

\begin{figure*}
  \includegraphics[width=\linewidth]{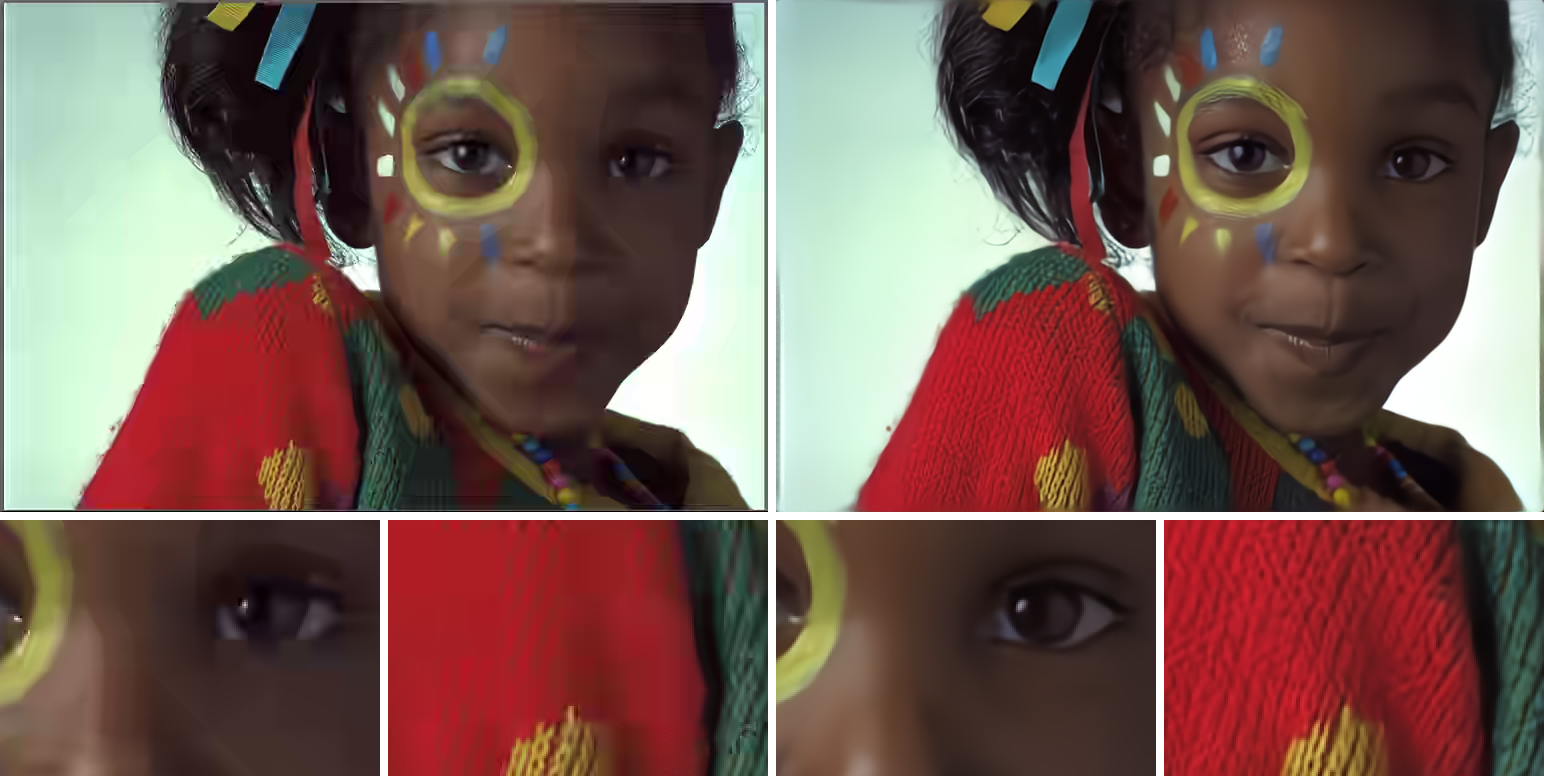}
  \caption{Reconstructions of kodim15 \citep{Kodak} compressed by BPG (left, 0.0738 bpp) and by a learned NTC model~\citep{MiSi20} optimized for MS-SSIM (right, 0.0713 bpp). The combination of NTC and an MS-SSIM loss function, which is designed to model texture masking effects in the human visual system, leads to significantly better texture retention in the sweater and fewer geometric distortions on the person's face.}
  \label{fig:kodim15}
\end{figure*}

While many authors have explored ANN-based compression in the context of existing, commercially viable image and video compression methods \citep{LaOs16,jiang2017end,li2018fully,jia2019content,liu2018cnn,hu2019progressive,HePfScRiSc19}, other authors begin with nonlinear transforms, and explore incorporating concepts traditionally used in linear TC, such as energy compaction \citep{cheng2019energy}, wavelets \citep{akyazi2019learning,yang2019deep}, or trellis coded quantization \citep{liu2020deep}. Lossless image compression based on learned entropy models has been explored by, e.g., \citet{mentzer2018conditional,mentzer2019practical}. Still others explore the intersection between learned image compression and other vision tasks such as content and semantic analysis \citep{li2018learning,campos2019content,akbari2019dsslic,luo2018deepsic}, inpainting \citep{baig2017learning}, super-resolution \citep{cao2020lossless}, quality enhancement \citep{lee2019hybrid}, or encryption \citep{duan2019efficient}.

Another topic of active research is the question of more “perceptual” image compression.
\citet{BaMiSiHwJo18} discuss optimization of NTC models for squared error vs. MS-SSIM.
\citet{ding2020comparison} provide a more in-depth discussion, with an even larger set of different perceptual distortion measures.
Interestingly, \citet{chen2019proxiqa} provide a method to optimize NTC models for non-differentiable perceptual metrics.
\citet{valenzise2018quality,cheng2019perceptual,ascenso2020learning} study the perceived image quality of learned image compression models optimized for metrics such as squared error and MS-SSIM with the help of human rating experiments.
Other authors have explored augmenting the fixed distortion measure with ANN-based losses that have shown visually convincing results in image generation tasks, most notably generative adversarial networks (GANs) \citep{santurkar2018generative,agustsson2019generative,tschannen2018deep,kudo2019gan,mentzer2020high}.
\citet{blau2019rethinking} formulate theoretical limits for the three-way trade-off between the rate, the reconstruction quality of an image compression method, as well as divergence measures between the source distribution and the marginal distribution of image reconstructions, the latter of which are related to adversarial losses.

While image compression dominates the literature on NTC, other applications have emerged as well, such as compression of point clouds \citep{quach2019learning,guarda2020deep}, volumetric data \citep{tang2020deep}, ANN features for tasks like large-scale image retrieval \citep{singh2020featcomp}, and compression of ANN parameters themselves \citep{oktay2019scalable}. Further reviews of the current state of the literature, specifically with respect to image and video compression, are given by \citet{ma2019image,liu2020deep}.

\section{Conclusion}
We have presented an overview of nonlinear transform coding, which heavily relies on ANNs as universal function approximators and stochastic optimization of the rate--distortion Lagrangian. We introduced the VECVQ algorithm, a stochastically optimized version of entropy-constrained VQ, as well as a novel method of $\lambda$-parameterization. Furthermore, we provide the first direct comparison of different $\lambda$-parameterizations for image compression models.

Most of the desirable properties of NTC stem from the use of stochastic RD optimization and ANNs. Given enough computing power, NTC models are quickly adaptible to arbitrary data sources, including domain-specific imagery (e.g., medical, astronomical) or new modalities of multimedia, since many parameters of the system can be found using end-to-end optimization rather than manual experimentation. This can lower prototyping times from years to weeks.

NTC models can also be directly optimized for any differentiable distortion measure (or, more generally, distortion loss). \Cref{fig:kodim15} illustrates the visual difference of a model optimized for MS-SSIM vs. squared error, which has long been noted as perceptually flawed \citep{Gi93}. This adaptability of NTC will dovetail with the development of better and more general perceptual losses, of which hybrid adversarial losses are an example \citep{santurkar2018generative,agustsson2019generative,tschannen2018deep,kudo2019gan,mentzer2020high}.

With the rapidly increasing availability of parallelized computation, we believe NTC will fundamentally change the landscape of practical data compression.

\section*{Acknowledgements}
The authors would like to thank the anonymous reviewers, as well as Lucas Theis and Aaron Wagner for their helpful comments on the draft and insightful discussions.

\printbibliography

\clearpage
\section{Supplemental materials}

\subsection{Local properties of nonlinear transforms}
\label{sec:local_properties}
In this section, we appeal to high-rate quantization theory \citep{Gersho79,GrNe98} to claim that an optimal nonlinear analysis transform has an inverse Jacobian whose columns tend towards orthogonality, at least in the limit of high rate.  In this sense, at high rates, the optimal nonlinear analysis transform is locally orthogonal, approximately.  Similarly the optimal nonlinear synthesis transform has a Jacobian whose columns tend toward orthogonality.  To simplify the technical argument, we first assume that the dimension $M$ of the latent space is the same as the source dimension, and that the transforms are at least locally invertible.  We also assume here that the distortion measure is the squared error.  Later we comment on how to lift these assumptions.

Let $\bm x$, ${\bm y}$, ${\bm k}$, and $\bm{\hat x}$ denote source, latent, quantized latent, and reproduction vectors, respectively, all with dimension $M$.  Let ${\bm y}=g_a(\bm x)$ denote the analysis transform, let ${\bm k}=\lfloor{\bm y}\rceil$ denote uniform scalar quantization of ${\bm y}$ by component-wise rounding, and let $\bm{\hat x}=g_s({\bm k})$ denote the synthesis transform.  Let $q({\bm k})$ denote a probability mass function on the integer vectors ${\bm k}$ (typically fully-factorized across the components of ${\bm k}$).  Assume $g_a$, $g_s$, and $q$ are optimal for some very large Lagrange multiplier $\lambda$.  That is, they minimize $L = \lambda D + R$, where
\begin{align}
D &= \E_{\bm x\sim p} \, \Bigl\|\bm x-g_s\bigl(\lfloor g_a(\bm x)\rceil\bigr)\Bigr\|^2, \\
R &= \E_{\bm x\sim p} \, \Bigl[-\log q\bigl(\lfloor g_a(\bm x)\rceil\bigr)\Bigr],
\end{align}
and $p(\bm x)$ is the density of $\bm x$.

Further, let $S_{\bm k}=g_a^{-1}({\bm k}+[-0.5,0.5)^M)$ be the quantization cell containing all source vectors $\bm x$ that map to ${\bm k}=\lfloor g_a(\bm x)\rceil$, let $\bm x_{\bm k}=g_a^{-1}({\bm k})\in S_{\bm k}$ be the source vector that maps exactly to ${\bm k}=g_a(\bm x)$, and let $\bm{\hat x}_{\bm k}$ be the centroid of $S_{\bm k}$.  Note that $\bm{\hat x}_{\bm k}=g_s({\bm k})$ since the synthesis transform $g_s$ is optimal.

Since $\lambda$ is large, we are in the high-rate regime.  In this regime, the cells $S_{\bm k}$ are small.  Thus, for all $\bm x\in S_{\bm k}$, $g_a(\bm x)\approx g_a(\bm x_{\bm k})+\nabla g_a(\bm x_{\bm k}) \cdot (\bm x-\bm x_{\bm k})$, where $\nabla g_a$ is the $M\times M$ Jacobian of $g_a$.  In other words, ${\bm y}\approx{\bm k}+\nabla g_a(\bm x_{\bm k}) \cdot (\bm x-\bm x_{\bm k})$, or $\bm x\approx\bm x_{\bm k}+\nabla g_a(\bm x_{\bm k})^{-1} \cdot ({\bm y}-{\bm k})$.  Thus $S_{\bm k}\approx\bm x_{\bm k}+\nabla g_a(\bm x_{\bm k})^{-1} \cdot [-0.5,0.5)^M$ is the image of the hypercube $[-0.5,0.5)^M$ under the linear map $\nabla g_a(\bm x_{\bm k})^{-1}$, centered on $\bm x_{\bm k}$.  In other words, $S_{\bm k}$ is a hyper ($M$-dimensional) parallelepiped, whose edge vectors are given by the columns of $\nabla g_a(\bm x_{\bm k})^{-1}$, and whose volume is $\det(\nabla g_a(\bm x_{\bm k})^{-1})$.

Now, as in \cite[eq. (25)]{GrNe98}, approximate the partial distortion in cell $S_{\bm k}$ as
\begin{align}
D_{\bm k}
&= \frac 1 M \int_{S_{\bm k}} \| \bm x - \bm{\hat x}_{\bm k} \|^2 p(\bm x) \D\bm x \notag \\
&\approx p(\bm{\hat x}_{\bm k}) m(S_{\bm k}) V(S_{\bm k})^{1+2/M},
\end{align}
where $V(S_{\bm k})$ is the volume of $S_{\bm k}$ and $m(S_{\bm k})$ is its \emph{normalized second moment of inertia} \citep{Gersho79,GrNe98}, defined by
\begin{equation}
m(S_{\bm k})
= \frac 1 M \frac 1 {V(S_{\bm k})^{1+2/M}}
\int_{S_{\bm k}} \| \bm x - \bm{\hat x}_{\bm k} \|^2 \D\bm x.
\end{equation}
Note that $m(S_{\bm k})$ is normalized such that it is a function of only the shape of $S_{\bm k}$, not its scale \citep{Gersho79}, i.e.,
\begin{align}
m(\alpha S_{\bm k}) &= \frac 1 M \frac 1 {V(\alpha S_{\bm k})^{1+2/M}}
    \int_{S_{\bm k}}\|\alpha\bm x-\alpha\bm{\hat x}_{\bm k}\|^2 \D(\alpha \bm x) \notag \\
&= \frac{1}{M}\frac{\alpha^{M+2}}{\alpha^{M(1+2/M)}V(S_{\bm k})^{1+2/M}} \int_{S_{\bm k}}\|\bm x-\bm{\hat x}_{\bm k}\|^2 \D\bm x \notag \\
&= m(S_{\bm k}).
\end{align}
It is intuitively obvious that among hyper parallelepipeds, $m(S_{\bm k})$ is minimized by a hyper cube, for which the value is $1/12$.

The overall distortion can thus be approximated as
\begin{align}
D = \sum_{\bm k} D_{\bm k}
&\approx \sum_{\bm k} p(\bm{\hat x}_{\bm k})m(S_{\bm k})V(S_{\bm k})^{1+2/M} \notag \\
&\approx \int p(\bm x) m(\bm x) V(\bm x)^{2/M} \D\bm x,
\end{align}
where $V(\bm x)$ is the volume of the cell containing $\bm x$, and likewise $m(\bm x)$ is the normalized second moment of inertia of the cell containing $\bm x$.

The overall rate can also be approximated, as
\begin{align}
R &\approx -\frac 1 M \sum_{\bm k} p(\bm{\hat x}_{\bm k}) V(S_{\bm k}) \log\bigl(q(\bm{\hat x}_{\bm k})V(S_{\bm k})\bigr) \notag \\
&\approx -\frac 1 M \int p(\bm x) \log\bigl(q(\bm x)V(\bm x)\bigr) \D\bm x \\
&= \frac 1 M \Bigl(\KL p q + h[p]\Bigr) - \frac 1 M \int p(\bm x)\log V(\bm x) \D\bm x, \notag
\end{align}
where $q(\bm x)$ denotes the density such that $q(\bm x)=q(\bm k)/V(S_{\bm k})$ for all $\bm x \in S_{\bm k}$, $\KL p q$ is the Kullback--Leibler divergence between $p$ and $q$, and $h[p]$ is the differential entropy of $p$.

Putting these together,
\begin{align}
L &\approx \E_{\bm x \sim p}\bigr[\lambda m(\bm x)V(\bm x)^{2/M} - \tfrac 1 2 \log V(\bm x)^{2/M}\bigl] 
\label{eqn:Lagrangian_high_rate_left}
\\
&\phantom{\approx} + \tfrac 1 M \KL p q + \const.
\label{eqn:Lagrangian_high_rate_right}
\end{align}
We now claim that a linear orthogonal transform $g_a$ (and its corresponding $q$) will nearly minimize the two terms (\ref{eqn:Lagrangian_high_rate_left}) and (\ref{eqn:Lagrangian_high_rate_right}).  To see this, observe that if we could minimize them independently over $q$, $m$, and $V$, we would choose $q(x)=p(x)$, $m(x)=1/12$, and $V(x)=((2\ln 2)\lambda m(x))^{-2/M}=((\ln 2)\lambda/6)^{-2/M}$.  The corresponding transform $g_a$ would be a linear orthogonal transform, which has hyper cubical quantization cells all the same size.
Unfortunately, this is generally not a viable solution for transform codes, since $q$ is typically constrained to be factorized along the components of $g_a$.  Hence, $\KL p q$ will generally not be zero (unless $p$ itself can be factorized along the components of $g_a$, for example if $p$ is Gaussian and the orthonormal transform is oriented along the principal axes of the Gaussian).  Thus when $q$ is so constrained, there is generally a trade-off between minimizing the two terms.

In the limit of high rate, as $\lambda$ becomes large, (\ref{eqn:Lagrangian_high_rate_left}) dominates (\ref{eqn:Lagrangian_high_rate_right}).  The former is minimized by the choices given above, corresponding to an orthogonal transform with hyper cubical quantization cells all the same size. Thus, orthogonal transforms are near-optimal in the high-rate case. (This is a special case of the usual high-rate quantization theory result that the optimal entropy-constrained vector quantizer
is formed by tesselating polytopes with minimum normalized moment of inertia \citep{Gersho79}.)  

For realistic bit rates, of course, (\ref{eqn:Lagrangian_high_rate_right}) matters.  For a given transform $g_a$ (and hence $m$ and $V$), the optimal $q=q(g_a)$ will minimize $\KL p q$ subject to the constraint that $q$ is factorizable over the components of $g_a$.  Since both terms depend on $g_a$, they can no longer be independently minimized.  Thus $m(\bm x)$ may no longer be minimal, meaning that the quantization cells may no longer be cubical.  Furthermore, $V(\bm x)$ may no longer be constant, meaning that the codewords may no longer have a uniform density.  As $\lambda$ changes the target bit rate, so it will change the character of the transform, as the quantization cells evolve from hyper cubes at high rates to other shapes at lower rates.  Through such evolution, however, (\ref{eqn:Lagrangian_high_rate_left}) always biases the solution toward quantization cells that have as small normalized moment of inertia as possible, i.e., towards hyper cubes.  It is for this reason that the local Jacobian and inverse Jacobian matrices of an optimal nonlinear transform $g_a$ tend towards orthogonality, especially at high rates, even as globally the transform becomes warped.

\enlargethispage{-3.2in}

Now let us comment on how one might lift some of our assumptions.  In practice, of course, the latent dimension $M$ is frequently chosen lower than the source dimension, say $M_0$.  In this case, the analysis transform $g_a^{(M)}$ cannot be invertible, and its Jacobian is not square and thus has no inverse.   However, suppose that $g_a^{(M_0)}$ is the analysis transform optimal for the case where the latent dimension is the same as the source dimension, at a particular level of distortion $D$, as described above.  Further suppose all but $M$ of the latent variables of $g_a^{(M_0)}$ consistently quantize to zero.  Then the source distribution must be largely confined to an $M$-dimensional manifold, at least within distortion $D$.  Letting $g_a^{(M)}$ be the analysis transform obtained from $g_a^{(M_0)}$ by removing the latent variables that consistently quantize to zero, it is clear that the Jacobian of $g_a^{(M)}$ is the same as the Jacobian of $g_a^{(M_0)}$, with some rows removed.  More generally, as long as $M$ is not chosen below the dimension essential for reproducing the source at a given level of distortion, the inverse image of a quantization cube ${\bm k}+[-0.5,0.5)^M$ in the latent space is a quantization cell $S_{\bm k}$ in the $M$-dimensional manifold.  Within the manifold, the normalized moment of inertia of $S_{\bm k}$ should still tend to be minimized, and hence the $M$ rows of the Jacobian of $g_a^{(M)}$ should still tend to be orthogonal, using arguments similar to those above.

The other assumption that one might lift is the assumption that the distortion measure is the squared error.  The basic conclusions above should hold for any doubly differentiable distortion loss (e.g., most any perceptual distortion function, as well as hybrid adversarial losses), after a local coordinate transformation.  To proceed, suppose for all $\bm x$, $d(\bm x,\bm{\tilde x})$ as a function of $\bm{\tilde x}$ is twice continuously differentiable in $\bm{\tilde x}$ and that it is minimized when $\bm{\tilde x}=\bm x$. Then
\begin{equation}
    d(\bm x,\bm{\tilde x})\approx d(\bm x,\bm x) + (\bm{\tilde x}-\bm x)^\T H(\bm x)(\bm{\tilde x}-\bm x)
\end{equation}
is the Taylor approximation of $d(\bm x,\bm{\tilde x})$ in $\bm{\tilde x}$ around $\bm x$, where $H(\bm x)$ is the symmetric positive definite Hessian of $d(\bm x,\cdot)$ at $\bm x$.  Assume the Hessian is constant in the vicinity of $\bm x$, namely $H(\bm x)=H$, and that it can be factored into its square roots as
$H=R^TR$.  Then (with the inessential assumption that $d(\bm x,\bm x)=0$), the distortion in the vicinity of $\bm x$ can be expressed as the squared error
\begin{equation}
    d(\bm x,\bm{\tilde x})\approx ||R(\bm{\tilde x}-\bm x)||^2.
\end{equation}
Thus, at high rates, the conclusions of this section will hold in the vicinity of $\bm x$, after a local coordinate transformation by $R$, the square root of the Hessian of $d(\bm x,\cdot)$ at $\bm x$.  (When $d(\bm x, \bm{\tilde x})$ as a function of $\bm{\tilde x}$ is not minimized by $\bm{\tilde x}=\bm x$, as might be the case for hybrid adversarial losses, an additional local coordinate transformation may be necessary.)  For perceptual distortion measures whose Hessians are constant over the entire space, such a trick can be used for traditional linear transforms.  However, in the usual case where the Hessian varies over the space (e.g., to reflect perceptual masking effects), nonlinear transforms are uniquely able to build the perceptual measure directly into the optimal transform.

\end{document}